\documentclass{emulateapj}
\usepackage{natbib}
\usepackage{apjfonts}

\usepackage{amsmath}

\begin{document}
\title{A Method for Measuring (Slopes of) the Mass Profiles of Dwarf Spheroidal Galaxies}
\shorttitle{A Method for Measuring (Slopes of) the Mass Profiles in dSph Galaxies}
\author{Matthew G. Walker\altaffilmark{1,2,3} \& Jorge Pe\~narrubia\altaffilmark{2}}
\email{mwalker@cfa.harvard.edu}
\altaffiltext{1}{Harvard-Smithsonian Center for Astrophysics, 60 Garden St., Cambridge, MA 02138}
\altaffiltext{2}{Institute of Astronomy, University of Cambridge, Madingley Road, Cambridge CB30HA}
\altaffiltext{3}{Hubble Fellow}
\journalinfo{Accepted for publication in \textit{The Astrophysical Journal}}

\begin{abstract} 
We introduce a method for measuring the slopes of mass profiles within dwarf spheroidal (dSph) galaxies directly from stellar spectroscopic data and without adopting a dark matter halo model.  Our method combines two recent results: 1) spherically symmetric, equilibrium Jeans models imply that the product of halflight radius and (squared) stellar velocity dispersion provides an estimate of the mass enclosed within the halflight radius of a dSph stellar component, and 2) some dSphs have chemo-dynamically distinct stellar \textit{sub}components that independently trace the same gravitational potential.  We devise a statistical method that uses measurements of stellar positions, velocities and spectral indices to distinguish two dSph stellar subcomponents and to estimate their individual halflight radii and velocity dispersions.  For a dSph with two detected stellar subcomponents, we obtain estimates of masses enclosed at two discrete points in the same mass profile, immediately defining a slope.  Applied to published spectroscopic data, our method distinguishes stellar subcomponents in the Fornax and Sculptor dSphs, for which we measure slopes $\Gamma\equiv \Delta \log M / \Delta \log r=2.61_{-0.37}^{+0.43}$ and $\Gamma=2.95_{-0.39}^{+0.51}$, respectively.  These values are consistent with `cores' of constant density within the central few-hundred parsecs of each galaxy and rule out `cuspy' Navarro-Frenk-White (NFW) profiles ($d\log M/d\log r \leq 2$ at all radii) with significance $\ga 96\%$ and $\ga 99\%$, respectively.  Tests with synthetic data indicate that our method tends systematically to overestimate the mass of the inner stellar subcomponent to a greater degree than that of the outer stellar subcomponent, and therefore to underestimate the slope $\Gamma$ (implying that the stated NFW exclusion levels are conservative). 
\end{abstract}


\section{Introduction}
\label{sec:intro}

Cold dark matter (CDM) halos produced in collisionless cosmological $N$-body simulations follow a nearly universal mass-density profile that diverges toward the center as $\lim_{r\rightarrow 0}\rho(r)\propto r^{-\gamma}$ with $\gamma\ga 1$, forming a so-called `cusp' (\citealt{dubinski91}, \citealt{navarro96,navarro97} (`NFW' hereafter), \citealt{moore98}, \citealt{klypin01}, \citealt{diemand05b}, \citealt{springel08}).  Many observations aim to test this scenario by using the measured motions of dynamical tracers in individual galaxies to constrain slopes of the underlying dark matter density profiles (e.g., \citealt{moore94,flores94,deblok97,salucci00,mcgaugh01,simon05}, \citealt[and references therein]{deblok10}).  However, comparisons to cosmological models tend to be inconclusive for the simple reason that while most cosmological N-body simulations consider only dark matter particles, one observes only baryons.  Baryons complicate not only the measurement of a dark matter density profile but also its interpretation within the context of the CDM paradigm.  Complicating the measurement is the fact that any uncertainty (e.g., uncertain stellar mass-to-light ratios) in the baryonic mass profile propagates to the inferred dark matter profile, as the latter is merely the difference between dynamical and baryonic mass profiles.  Complicating the interpretation of even a clean measurement is the possibility that various poorly-understood dynamical processes involving baryons might alter the original structure of a dark matter halo (e.g., \citealt{blumenthal86,navarro96b,elzant01,gnedin04,tonini06,governato10,pontzen11}). 

One mitigates both of these complications at once by considering the Milky Way's (MW's) dwarf spheroidal (dSph) satellites, which have the smallest sizes ($\sim 10^{2-3}$ pc), smallest baryonic masses ($L_{V}\sim 10^{3-7} L_{\odot}$) and the largest dynamical mass-to-light ratios ($[M/L_V]/[M/L_V]_{\odot}\ga 10$) of any observed galaxies (\citealt{aaronson83}, \citealt[][and references therein]{mateo98}, \citealt{gilmore07}).  Dark matter dominates even at the centers of dSph gravitational potential wells, implying that uncertainties in baryonic mass profiles have negligible impact on inferred dark matter profiles.  Furthermore, dynamical processes that invoke baryon physics to alter the central structure of dark matter halos are subject to strong constraints in dSphs, where one finds the smallest baryon densities and infers the largest dark matter densities of any galaxy type \citep{pryor90}.  

Recent work identifies several mechanisms that in principle are capable of altering the central structure---i.e., of transforming primal `cusps' into `cores' of constant density---of cold dark matter halos on dSph-like scales.  Such mechanisms typically invoke either the dynamical coupling of the dark matter to rapid baryonic outflows (e.g., \citealt{read05,mashchenko06,mashchenko08,desouza11}) or the transfer of energy and/or angular momentum to the dark matter from massive infalling objects (e.g., \citealt{sanchez06,goerdt06,goerdt10,cole11}).  While both channels for baryon-driven core creation are physically plausible, their application to real dSphs must eventually satisfy a large body of observational constraints.  It is therefore instructive to consider recent high-resolution hydrodynamical simulations by \citet{sawala10} and \citet{parry11}.  These cosmological simulations broadly reproduce the observed distributions of luminosities, metallicities and stellar kinematics exhibited by Local Group dSphs.  Both groups conclude that the baryon-physical processes driving the formation and evolution of their simulated dSphs leave the cuspy central structure of dSph CDM halos intact.  Taking these results at face value, it seems then that the Local Group dSphs represent the most pristine dark matter halos to which we have observational access.  Measurements of the slopes (i.e. `cusp' versus `core') of dSph mass profiles can therefore provide a uniquely direct test of structure formation within the CDM paradigm.

Pressure-supported stellar components provide the only available kinematic tracers in dSphs, but thus far stellar kinematic data have figured only indirectly in core/cusp investigations.  For example, \citet{kleyna03} detect kinematically cold stellar substructure in the Ursa Minor dSph and argue that its survival against tidal disruption is more likely in a cored as opposed to a cusped host potential.  \citet{sanchez06} and \citet{goerdt06} argue that the wide spatial distribution of the five globular clusters in the Fornax dSph again favors a cored host potential, as dynamical friction within a centrally cusped potential would have caused the clusters to sink to Fornax's center in less than a Hubble time (unless those clusters had much wider orbits initially).  On the other hand, \citet{penarrubia10} argue that the mass--size relation traced by the Milky Way's dSph population favors evolutionary scenarios that invoke cusped as opposed to cored halos\footnote{This result is particularly sensitive to the masses inferred for the Milky Way's `ultrafaint' satellites.  \citet{mcconnachie10} have recently shown that the small velocity dispersions observed for many of these systems can receive significant contributions from binary orbital motions, a conclusion supported by the recent direct detection of resolved binary motions in the Bo\"otes I satellite \citep{koposov11}.  Downward revision of the intrinsic velocity dispersions (and hence masses) of several of the smallest ultrafaint dSphs could lead to a size/mass relation for Milky Way satellites that favors cored over cusped dark matter halos (see Figure 11 of \citealt{penarrubia10}).}.

In contrast to the studies mentioned above, here we devise a method for measuring the slopes of dSph mass profiles directly from stellar spectroscopic data.  We proceed by combining two recent results.  First, for a spherically symmetric dSph in dynamic equilibrium, the product of halflight radius and (squared) velocity dispersion provides an estimate of the mass enclosed within the halflight radius \citep{walker09d,wolf10}.  Second, some dSphs contain at least two chemo-dynamically distinct stellar populations \citep{tolstoy04,battaglia06,battaglia11}, each presumably tracing the same dark matter potential.  Here we formulate a mathematical model that uses measurements of stellar positions, velocities and spectral indices to distinguish two dSph stellar subcomponents and to estimate their individual halflight radii and velocity dispersions.  For a dSph with two detected stellar subcomponents, we obtain estimates of masses enclosed at two discrete points in the same mass profile.  Two points define a slope.

\subsection{Stellar Kinematics with Two Numbers}
\label{subsec:twonumbers}

In principle the Collisionless Boltzmann Equation (CBE, Equation 4.6 of \citealt{bt08}) relates the 6-dimensional phase-space distribution function, $f(\vec{r},\vec{v})$, of a tracer component to the underlying gravitational potential, thereby governing the joint distribution of stellar positions and velocities for a pressure-supported galaxy in dynamic equilibrium.  In practice the available dSph data provide information in only three dimensions---two spatial dimensions orthogonal to the line of sight and one velocity dimension along the line of sight.  Implementation of the CBE with dSph data then requires transformations between 6D and 3D (or 2D with spherical symmetry) phase-space distributions (e.g., \citealt{wilkinson02}), often at significant computational expense.

\begin{figure}
  \epsscale{1.1}
  \plotone{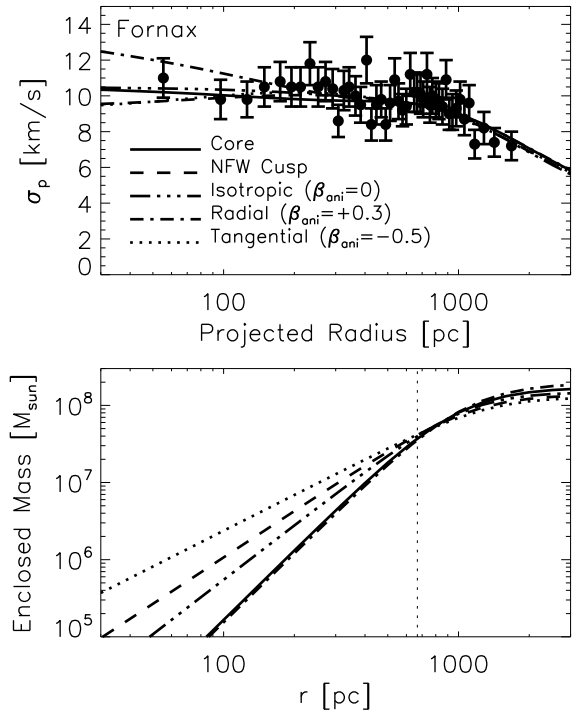}
  \caption{\scriptsize \textit{Top:} Projected stellar velocity dispersion profile for the Fornax dSph adopted from \citet{walker09d}.  Overlaid are spherical Jeans models that assume either a cored dark matter halo, an NFW dark matter halo, or if one lets the shape of the dark matter halo vary freely, velocity distributions that are either isotropic, radially anisotropic, or tangentially anisotropic.  \textit{Bottom:} Enclosed-mass profiles corresponding to the same models.  The vertical dotted line indicates Fornax's projected halflight radius \citep{ih95}, where the simple estimator specified by Equation \ref{eq:walker} gives $M(r_h)=[5.3\pm 0.9] \times 10^7M_{\odot}$, in agreement with the value common to the various successful Jeans models.}
  \label{fig:for_profiles}
\end{figure}

Many dSph kinematic studies (e.g., \citealt{wilkinson04,strigari06,strigari08,koch07,lokas09,walker09d,battaglia08,battaglia11}) rely instead on the Jeans equations, obtained by integrating the CBE over velocity space.  The spherically symmetric Jeans equation specifies the mass profile $M(r)$---including the contribution from any dark matter component---in terms of the stellar density profile, $\nu(r)$, and stellar velocity dispersion profile, $\bar{v^2}(r)$ \citep{bt08}:
\begin{equation}
  \frac{1}{\nu}\frac{d}{dr}(\nu \bar{v_r^2})+\frac{2}{r}(\bar{v_r^2}-\bar{v_{\theta}^2})=-\frac{GM(r)}{r^2},
  \label{eq:jeans}
\end{equation}
where $\bar{v^2_r}$ and $\bar{v^2_{\theta}}$ are components of the velocity dispersion in radial and tangential directions, respectively.  Confinement of dSph stellar velocity data to the component along the line of sight leaves the velocity anisotropy---usually quantified by the ratio $\beta_{\mathrm{ani}}(r)\equiv 1-\bar{v^2_{\theta}}(r)/\bar{v^2_r}(r)$---poorly constrained, ultimately precluding model-independent constraints on the mass profile in analyses based on Equation \ref{eq:jeans}.  For example, the top panel of Figure \ref{fig:for_profiles} demonstrates that the projected velocity dispersion profile observed for the Fornax dSph can be fit equally well by Jeans models that assume either cored or NFW-cusped dark matter halos, or if the shape of the dark matter halo is unspecified, by models that assume the velocity distribution is either isotropic, radially anisotropic or tangentially anisotropic.  

The bottom panel of Figure \ref{fig:for_profiles} demonstrates that despite this well-known degeneracy between mass and anisotropy, the various successful Jeans models tend to have the same mass enclosed within approximately the dSph halflight radius (e.g., \citealt{strigari07,penarrubia08a,walker09d,wolf10,amorisco11}).  Furthermore, given the general flatness of dSph velocity dispersion profiles \citep{walker07b}, the value of this mass relates simply to the product of velocity dispersion and the halflight radius of the adopted surface brightness model.  Assuming that the stellar component follows a Plummer profile ($\nu(r)\propto [1+r^2/r_h^2]^{-5/2}$) and has an isotropic ($\bar{v^2_r}=\bar{v^2_{\theta}}$) velocity distribution with constant dispersion, \citet{walker09d} derive from Equation \ref{eq:jeans} the simple estimator
\begin{equation}
  M(r_h)\approx \frac{5r_h\sigma^2_V}{2G},
  \label{eq:walker}
\end{equation}
where $r_h$ is the projected halflight radius and $\sigma^2_V$ is the square of the velocity dispersion.  Equation \ref{eq:walker} provides estimates of $M(r_h)$ that agree well with formal constraints from searches of many-dimensional parameter spaces often invoked in Jeans models (see Figure 3 of \citealt{walker09d}).  This agreement follows from the fact that for flat velocity dispersion profiles and the adoption of a Plummer surface brightness profile, the two observables $r_h$ and $\sigma^2_V$ already contain all of the empirical information that enters Equation \ref{eq:jeans}.  

Thus Equation \ref{eq:walker} provides reasonably model-independent estimates of masses enclosed within dSph projected halflight radii (see section \ref{subsubsec:masses} for an examination of bias).  \citet{wolf10} show that an alternative estimator, which can be expressed as $M(\frac{4}{3}r_h)\approx 4r_h\sigma^2_V/G$, gives model-independent estimates of dSph masses at slightly larger radii (approximately the deprojected halflight radius) where good-fitting mass profiles intersect with slightly less scatter.  For our purposes here, all estimators of the form $M(\kappa r_h)\propto r_h\sigma^2_V$, where $\kappa$ is some constant, are equivalent.

\subsection{Distinct Stellar Subcomponents in dSphs}
\label{subsec:twocomponents}

The ability to make a robust estimate of the mass enclosed within the halflight radius of a dSph stellar population takes on additional significance given recent discoveries that at least some dSphs have multiple stellar populations, each of which can serve as an independent tracer of the underlying gravitational potential.  Using VLT/FLAMES spectroscopy of $\sim 400$ stars in the Sculptor dSph, \citet{tolstoy04} find evidence for two ancient stellar subcomponents that follow distinct distributions in position, velocity and metallicity.  Imposing a metallicity cut to divide their spectroscopic sample into relatively `metal-rich' and `metal-poor' subcomponents, \citet{tolstoy04} find that the metal-rich subcomponent is more centrally concentrated and kinematically colder than the metal-poor subcomponent.  \citet{battaglia06} and \citet{battaglia11} report similar discoveries for the Fornax and Sextans dSphs based on VLT/FLAMES samples of $\sim 550$ and $\sim 200$ member stars, respectively---in both galaxies a relatively metal-rich subcomponent is again more centrally concentrated and has smaller velocity dispersion than a metal-poorer, kinematically hotter, more extended subcomponent.  \citet{ibata06} report similar phenomenology based on Keck/Deimos observations of 44 members of the CVnI dSph (although follow-up Keck/Deimos observations by \citealt{simon07} do not reproduce this result).  Several other dSphs, including Carina, Tucana and some M31 satellites display evidence for distinct stellar subcomponents in the form of differing spatial distributions of blue- and red-horizontal branch stars \citep{harbeck01,bellazzini01}, but these separations have not yet been linked to stellar kinematics.

There are two previously-published efforts to model dSphs as superpositions of two dynamically independent stellar subcomponents tracing the same dark matter potential.  \citet{mcconnachie07} find that the spatial and velocity distributions of several dSphs, including satellites of both the Milky Way and M31, are compatible with the presence of dynamically distinct stellar subcomponents embedded within cuspy NFW halos.  \citet{battaglia08} apply more general dark matter halo models, considering cored as well as cuspy NFW halos.  \citet{battaglia08} find that while both types of halos can plausibly host the two stellar subcomponents they detect in Sculptor, the cored halo gives a better fit to the falling velocity dispersion profile they measure for Sculptor's metal-rich subcomponent.  

In contrast to these previous studies, here we do not adopt a dark matter halo model.  Nor do we use rigid color/metallicity cuts to separate metal-rich and metal-poor subcomponents (c.f. \citealt{battaglia08}).  Instead we devise a statistical technique for the purpose of separating and quantifying the radius, metallicity and velocity distributions followed by two distinct dSph stellar subcomponents that independently trace the same gravitational potential.  Where we can resolve a dSph into two distinct stellar subcomponents and measure the halflight radii and velocity dispersions of both subcomponents, we effectively resolve two discrete points in a mass profile dominated by dark matter.  The slope then follows directly.

\section{Data}
\label{sec:data}

We use the spectroscopic data published by \citet[W09 hereafter]{walker09a} for three of the Milky Way's `classical' dSph satellites: Carina, Fornax and Sculptor.  These data were compiled over five years (2004--2008) of observations with the Magellan/Clay (6.5m) telescope and Michigan/MIKE Fiber Spectrograph (MMFS, PI: Mario Mateo, Co-I: Ed Olszewski) at Las Campanas Observatory, Chile.  MMFS spectra cover the range $5140$ \AA\ -- $5180$ \AA\ at high resolution ($R \sim 20000$), including the prominent magnesium triplet (MgT) absorption feature; see \citet[W07 hereafter]{walker07a} and W09 for details of observations and data reduction.  The resulting data set includes line-of-sight velocities and spectral indices measured individually for $\sim 6000$ red giant candidates in the three dSphs.  For the present study we exclude the $\sim 500$ stars for which W09 measure a velocity but do not report a value for the magnesium index (due to insufficient signal).  Table \ref{tab:mmfs} gives  central coordinates, heliocentric distances and integrated luminosities \citep{mateo98} for these three galaxies and lists the numbers of stars (including foreground interlopers) in the MMFS spectroscopic samples used here.  

For three reasons we exclude from the present study the available MMFS data for the Sextans dSph.  First, the MMFS data set for Sextans is relatively small, containing velocities for just $\sim 950$ stars, of which only $\sim 450$ are Sextans members.  Second, the method we introduce below requires calibrated broad-band photometry (Section \ref{subsec:velocities}), and as reported by W09, poor observing conditions precluded accurate calibration of the photometry used to select Sextans targets.  Third, in order to compensate for this shortcoming, W09 report calibrated photometry for $\sim 430$ of their Sextans targets ($\sim 300$ members) using the catalog of \citet{lee03}.  However, this catalog covers only a few percent of the total surface area of Sextans, complicating the correction for sampling bias that we discuss in Section \ref{subsec:selection}.  Thus a comparable analysis of Sextans, which we leave for future work, would require more complete photometry and/or careful accounting for these deficiencies.

\begin{deluxetable*}{lrrrrrrrrrr}
  \tabletypesize{\scriptsize}
  \tablewidth{0pc}
  \tablecaption{Positions, luminosities and numbers of Magellan/MMFS-observed stars for southern classical dSphs\tablenotemark{*}}
  \tablehead{\\
    \colhead{dSph}&\colhead{RA (J2000)}&\colhead{Dec. (J2000)}&\colhead{Distance}&\colhead{$M_V$}&\colhead{$M_{V,HB}$\tablenotemark{**}}&\colhead{$N_{\mathrm{sample}}$}\\
    \colhead{}&\colhead{[hh:mm:ss]}&\colhead{[dd:mm:ss]}&\colhead{[kpc]}&\colhead{[mag]}&\colhead{[mag]}\\
  }
  \startdata
  \\
  Carina&06:41:37&$-$50:58:00&$101\pm 5$&$-9.3$&$20.9$&$1481$\\
  Fornax&02:39:59&$-$34:27:00&$138\pm 8$&$-13.2$&$21.3$&$2603$\\
  Sculptor&01:00:09&$-$33:42:30&$79\pm 4$&$-11.1$&$20.1$&$1497$\\
  \enddata
  \tablenotetext{*}{Central coordinates, distances and absolute magnitudes are adopted from the review of \citet{mateo98}.}
  \tablenotetext{**}{references for horizontal-branch magnitudes: Carina \citep{koch06}, Fornax and Sculptor \citep{ih95,battaglia08}}
  \label{tab:mmfs}
\end{deluxetable*}

\subsection{Velocities and Mg-Triplet Indices}
\label{subsec:velocities}
The MMFS velocity sample has a median measurement error of $\sim 2$ km s$^{-1}$, smaller than the internal stellar velocity dispersions ($\sigma_V \ga 6$ km s$^{-1}$) measured for these three dSphs.  Although their observations were designed primarily to measure velocities, W07 demonstrate that spectral indices---i.e., pseudo-equivalent widths of resolved Fe and Mg absorption lines---measured from MMFS spectra correlate with stellar atmospheric parameters such as effective temperature, surface gravity and metallicity (see Figures 19-20 of W07).  W07 combine indices measured separately for magnesium-triplet lines at $5167$ \AA\ and $5173$ \AA\ into a composite MgT index, denoted $\Sigma\mathrm{Mg}$, that is qualitatively similar to the composite calcium-triplet index, $\Sigma\mathrm{Ca}$, often used to infer stellar metallicities from near-infrared calcium-triplet spectra (e.g., \citealt{armandroff91,koch06,battaglia08b,starkenburg10}).  

Like the $\Sigma\mathrm{Ca}$ index, the $\Sigma\mathrm{Mg}$ index provides a measure of stellar-atmospheric metal abundance, provided that empirical calibrations adequately remove the dependences of Mg opacity on effective temperature and surface gravity.  For red giant stars that brighten and cool as they expand, these dependences translate (for stars of a given metallicity) into a nearly linear empirical relationship between $\Sigma\mathrm{Mg}$ and luminosity.  W09 (their Figure 3) quantify this relationship using MMFS observations of red giants in six globular clusters that span the metallicity range $-2.0 \la \mathrm{[Fe/H]} \la -0.5$ and have negligible (for our purposes) internal metallicity spreads.  Following \citet{rutledge97b} and \citet{koch06}, W09 simultaneously fit the data for each cluster with straight lines sharing a common slope, obtaining
\begin{equation}
  \Sigma\mathrm{Mg}=-(0.079\pm 0.002)(V-V_{\mathrm{HB}})+\Sigma\mathrm{Mg}',
  \label{eq:wprime}
\end{equation}
where $V-V_{\mathrm{HB}}$ is the offset in $V-$band luminosity from the horizontal branch.  The slope quantifies the dependence of opacity on effective temperature and surface gravity, using luminosity as a proxy.  The intercept, or `reduced' index $\Sigma\mathrm{Mg}'$---henceforth denoted $W'$---represents the value of $\Sigma\mathrm{Mg}$ that the star would have if it had the surface gravity and temperature of a horizontal branch star.  Then taking the empirical calibration given by Equation \ref{eq:wprime} at face value, red giants of similar metallicity should have similar $W'$.

Using the horizontal-branch magnitudes listed in Table \ref{tab:mmfs}, we apply Equation \ref{eq:wprime} to obtain reduced magnesium indices $W'$ for all dSph stars in the Magellan/MMFS sample.  Since we are concerned with stellar-atmospheric chemistry only as a diagnostic with which to distinguish stellar subcomponents independently of their velocity distributions, we do not require further calibration of $W'$ values onto an absolute metallicity scale.  In what follows we shall use $W'$ as an indicator of \textit{relative} metallicity and we shall refer to subcomponents distinguished by $W'$ as `metal-rich' and `metal-poor'.

\subsection{Membership}
\label{subsec:membership}

W07 (see their Figure 1) chose dSph stellar targets from selection boxes overlaid on red giant branches (RGBs) apparent from $V$, $I$ photometry of each system.  These selection regions include contamination from foreground Milky Way stars (typically late-type dwarfs) and background point sources (unresolved galaxies, quasars).  Fortunately, bona fide dSph members follow conspicuous distributions in velocity, spectral index and position, which helps to distinguish them from contaminant populations.  

The MMFS data published by W09 include for each star a probability of dSph membership that is derived from an expectation-maximization (EM) algorithm \citep{walker09b}, similar to algorithms previously used to determine membership in open clusters (e.g., \citealt{sanders71}).  The EM algorithm used by \citet{walker09b} adopts the foreground velocity distribution given by the Besan{\c c}on Milky Way model \citep{robin03} and assumes a single dSph stellar component.  Because the latter assumption obviously is at odds with the two-subcomponent models we shall consider here, we use W09's membership probabilities only to provide non-parametric estimates of position, velocity and Mg-index distributions followed by non-members (section \ref{subsec:foreground}).  The accuracy with which we recover the input member fraction in tests of our method (Section \ref{subsec:mcmcresults}) indicates that these estimates are not unduly influenced by the subtle differences between foreground probabilities derived from single and multi-component dSph models.  

\subsection{Spatial Selection Bias}
\label{subsec:selection}

Finally, in light of the fact that here we use the positions of stars in the MMFS samples to estimate halflight radii of the stellar populations to which they belong, we must consider the fact that the spatial distributions of MMFS-observed stars may differ from those of the populations from which they are drawn.  Each MMFS configuration is limited to $\leq 256$ targets within a $25'$ field of view, and therefore the distribution of stellar positions is not sampled randomly.  

\begin{figure*}
  \epsscale{1.}
  \plotone{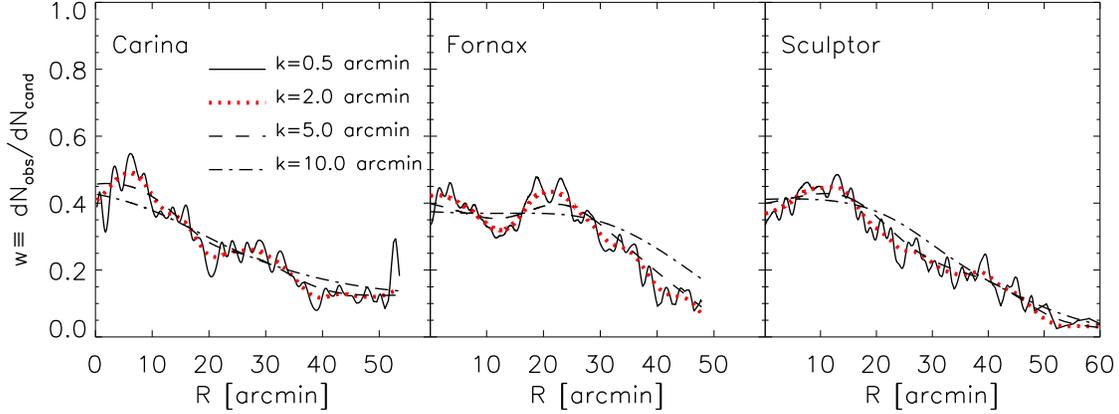}
  \caption{\scriptsize Spatial sampling bias in the Magellan dSph spectroscopic samples \citep{walker09a} adopted here.  Panels plot the observational selection probability, $w\equiv dN_{\mathrm{obs}}/dN_{\mathrm{cand}}$, as a function of projected radius, estimated via kernel smoothing (Equation \ref{eq:weight}).  Different linestyles correspond to different smoothing bandwidths.  For this study we adopt the $w(R)$ estimates that correspond to bandwidth $k_1=2$ arcmin (dotted red); our results and conclusions are not sensitive to this choice.  }
  \label{fig:weightbandwidth}
\end{figure*}

In order to compensate for selection effects due to the inevitable peculiarities of spectroscopic spatial sampling, we assign a selection probability to each star according to the local ratio of the number of observed stars, $dN_{\mathrm{obs}}$, to the number of target candidates, $dN_{\mathrm{cand}}$, selected according to W07's photometric criteria (the latter number includes both observed and unobserved stars).  We estimate this ratio as a function of projected radius by smoothing the data and target catalogs with Gaussian kernels:  
\begin{equation}
  \hat{w}(R)\equiv \frac{d\hat{N}_{\mathrm{obs}}(R)}{d\hat{N}_{\mathrm{cand}}(R)}\approx \frac{\displaystyle\sum_{i=1}^{N_{\mathrm{obs}}}\exp\biggl[-\frac{1}{2}\frac{(R_i-R)^2}{k_1^2}\biggr ]}{\displaystyle\sum_{i=1}^{N_{\mathrm{cand}}}\exp\biggl[-\frac{1}{2}\frac{(R_i-R)^2}{k_1^2}\biggr ]},  
  \label{eq:weight}
\end{equation}
where the hat ($\hat{\hspace{0.1in}}$) symbol denotes a quantity estimated via kernel smoothing.  Figure \ref{fig:weightbandwidth} displays $\hat{w}(R)$ curves for each dSph and for possible choices of bandwidth over the range $0.1 \leq k_1/\mathrm{arcmin})\leq 10$---these smoothing scales are smaller than the scale radius of the composite stellar component and larger than the (projected) mean free path of sampled stars.  For the present work we adopt $k_1=2$ arcmin (dotted red curves in Figure \ref{fig:weightbandwidth}), but we have confirmed that our results and conclusions are not qualitatively sensitive to this choice.

\section{Method}
\label{sec:method}

We model each dSph mathematically as the superposition of two chemo-dynamically distinct stellar subcomponents observed through foreground contamination.  For `metal-rich' and `metal-poor' stellar subcomponents we adopt simple parametric models to describe the distributions of projected radius, line-of-sight velocity and magnesium index (Sections \ref{subsec:rdist} - \ref{subsec:mgdist}).  For the foreground contamination we estimate these distributions non-parametrically by smoothing the spectroscopic data according to the published probability of (non-) membership for each star (Section \ref{subsec:foreground}).  We constrain model parameters using a standard Markov-Chain Monte Carlo algorithm (Section \ref{subsec:mcmc}).  The algorithm returns estimates of parameters including halflight radii and velocity dispersions for the two stellar subcomponents, which simultaneously provide estimates of the mass enclosed at the halflight radius of each subcomponent (Equation \ref{eq:walker}) as well as the slope of the mass profile:
\begin{equation}
  \Gamma\equiv \frac{\Delta \log M}{\Delta \log r}=\frac{\log [M(r_{\mathrm{h,2}})/M(r_{\mathrm{h,1}})]}{\log [r_{\mathrm{h,2}}/r_{\mathrm{h,1}}]}\approx 1+\frac{\log [\sigma^2_{V,2}/\sigma^2_{V,1}]}{\log [r_{\mathrm{h,2}}/r_{\mathrm{h,1}}]}.
  \label{eq:slope}
\end{equation}
The last expression on the right-hand side follows from Equation \ref{eq:walker} or, more generally, from any mass estimator of the form $M(\kappa r_h)\propto r_h\sigma^2$ where $\kappa$ is some constant (e.g. \citealt{penarrubia08a,walker09d,wolf10,amorisco11}), and makes explicit the fact that the only physical quantities relevant to our estimate of the slope $\Gamma$ are the sizes and velocity dispersions of the two stellar subcomponents.

\subsection{Likelihood Function}
\label{subsec:likelihood}

We require a mathematical model that will let us distinguish and quantify the properties of two independent stellar subcomponents in the same dSph.  Suppose $p_{1}(R,V,W')$ and $p_{2}(R,V,W')$ describe joint probability distributions\footnote{These probability densities are statistical distribution functions defined such that, for example, the fraction of subcomponent-1 stars that have position in the interval $R,R+dR$, velocity in the interval $V,V+dV$ and reduced magnesium index in the interval $W',W'+dW'$ is given by $p_{1}(R,V,W')dRdVdW'$.} of projected radius, line-of-sight velocity and reduced magnesium index for metal-rich and metal-poor dSph subcomponents, respectively.  Further suppose that $p_{\mathrm{MW}}(R,V,W')$ is the joint probability distribution followed by foreground Milky Way stars that satisfy the color/magnitude criteria used to select dSph targets.  Our data set $\{R_i,V_i,W'_i\}_{i=1}^{N_{\mathrm{sample}}}$ samples all three stellar populations.  Let the vector $\vec{S}$ represent a set of free parameters that specifies models for $p_{1}(R,V,W')$ and $p_{2}(R,V,W')$ as well as the fractions $f_{1}\equiv N_{1}/(N_{1}+N_{2}+N_{\mathrm{MW}})$ and $f_{2}\equiv N_{2}/(N_{1}+N_{2}+N_{\mathrm{MW}})$ of stars in metal-rich and metal-poor stellar subcomponents, respectively.  Recalling from Section \ref{subsec:selection} that $w(R)$ indicates the probability that an RGB candidate at radius $R$ is actually observed, then given a model specified by $\vec{S}$ the data set has likelihood 
\begin{eqnarray}
  L(\{R_i,V_i,W'_i\}_{i=1}^{N_{\mathrm{sample}}}|\vec{S})=\hspace{1.75in}\nonumber\\
  \displaystyle\prod_{i=1}^{N_{\mathrm{sample}}}\biggl [ f_{1}\frac{w(R_i)p_{1}(R_i,V_i,W'_i)}{\int\int\int w(R)p_{1}(R,V,W')dRdVdW'}\nonumber\\
+f_{2}\frac{w(R_i)p_{2}(R_i,V_i,W'_i)}{\int\int\int w(R)p_{2}(R,V,W')dRdVdW'}\nonumber\\
    +(1-f_{1}-f_{2})\frac{w(R_i)p_{\mathrm{MW}}(R_i,V_i,W'_i)}{\int\int\int w(R)p_{\mathrm{MW}}(R,V,W')dRdVdW'}\biggr ].
  \label{eq:likelihood}
\end{eqnarray}
The normalizing constant in the denominator of each term provides a weight that compensates for the radial sampling bias \citep{gill88,wang05,martinez11}.  

In the following subsections we specify mathematical models for $p_1$, $p_2$ and $p_{\mathrm{MW}}$ that let us evaluate the overall likelihood given by Equation \ref{eq:likelihood}.  In selecting from what is in principle an unlimited number of possible models, we opt for simplicity over elegance.  That is, in order to quantify distributions of positions, velocities and spectral indices we choose simple mathematical expressions that let us specify the likelihood function analytically and without introducing an unwieldy number of free parameters.  Since one can think of $p_1$, $p_2$ and $p_{\mathrm{MW}}$ as chemo-dynamical distribution functions, it is important to realize that the mathematical models that we adopt do not necessarily satisfy the Collisionless Boltzmann Equation and thus do not necessarily correspond to physical models.  It will therefore become necessary to explore the systematic errors introduced by the particular mathematical models that we adopt (see Section \ref{sec:tests}).

\subsection{Position Distributions for dSph stellar subcomponents}
\label{subsec:rdist}

dSph stellar surface density (number of stars per unit area) profiles typically are fit by \citet{plummer11}, \citet{king62} and/or \citet{sersic68}, profiles (e.g., \citealt{ih95,mcconnachie06,belokurov07,martin08}).  For simplicity we model the spatial distributions of member subcomponents with spherically symmetric Plummer profiles, for which the only free parameter is the projected halflight radius\footnote{The projected halflight radius is the radius of the circle enclosing half of the stars as observed in projection on the plane of the sky.}, $r_{h}$.  Then, for example, the metal-rich subcomponent (hereafter denoted with the subscript $1$) has stellar surface density given by  
\begin{equation}
  \Sigma_{1}(R)=\frac{N_{1}}{\pi r_{h,1}^2\bigl [1+R^2/r_{h,1}^2\bigr ]^2}.
  \label{eq:plummer1}
\end{equation}
(and similarly for the metal-poor subcomponent, hereafter denoted with the subscript $2$), where $N_1$ is the number of stars in subcomponent 1.  For a spherical system with stellar surface density profile $\Sigma(R)$, the 1-D probability distribution of projected radii is given by $p_R(R)=d/dR \bigl [\int_0^R \Sigma(S)SdS/\int_0^{\infty}\Sigma(S)SdS\bigr ]$, which for the Plummer profile becomes
\begin{equation}
  p_{R,1}(R)=\frac{2R/r^2_{h,1}}{\bigl (1+R^2/r^2_{h,1}\bigr )^2}
  \label{eq:plummerprob}
\end{equation}
(and similarly for the metal-poor subcomponent). 

\subsection{Velocity Distributions for dSph stellar subcomponents}
\label{subsec:veldist}

Again for simplicity, we assume that a given dSph stellar subcomponent has an intrinsic line-of-sight velocity distribution that is Gaussian and independent of radius.  Then the 1-D probability distribution of velocities is given by
\begin{equation}
  p_{V,1}(V,\alpha_*,\delta_*)=\frac{1}{\sqrt{2\pi (\sigma^2_{V,1}+\epsilon^2_V)}}\exp\biggl [-\frac{1}{2}\frac{(V-\langle V\rangle_{\alpha_*,\delta_*})^2}{\sigma^2_{V,1}+\epsilon^2_V}\biggr ]
  \label{eq:vprob}
\end{equation}
(and similarly for the metal-poor subcomponent), where $\sigma_{V,1}$ is the intrinsic velocity dispersion and $\epsilon_V$ is the velocity measurement error.  The dependence of the mean velocity $\langle V\rangle_{\alpha_*,\delta_*}$ on position follows from the fact that the velocity data are given in the heliocentric rest frame (HRF, i.e., the rest frame of an observer instantaneously comoving with the Sun).  Conversion to the dSph rest frame depends on relative motion between Sun and dSph, the transverse components of which are difficult to measure even from Hubble Space Telescope (HST) images separated by several years \citep{piatek02,piatek03,piatek06,piatek07}.  Furthermore, the line-of-sight velocity offset between HRF and the dSph rest frame varies systematically along the vector of any transverse motion between Sun and dSph, as this motion projects more strongly along lines of sight that are farther from the dSph center.   

In order to account for this `perspective effect' (e.g., \citealt{feast61,vandermarel02}) we follow \citet{kaplinghat08} and \citet{walker08} in considering a dSph at heliocentric distance $D$ with central equatorial coordinates $(\alpha_D,\delta_D)$, HRF line-of-sight velocity $V_D$ and HRF proper motion $(\mu_{\alpha},\mu_{\delta})$.  At the location of a star at $(\alpha_*, \delta_*)$ the dSph's systemic HRF velocity along the line of sight is (see Appendix of \citealt{walker08} for a derivation)
\begin{eqnarray}
  \langle V\rangle_{\alpha_*,\delta_*}=\hspace{2.5in}\nonumber\\
  \cos\delta_*\sin\alpha_*\bigl(V_D\cos\delta_D\sin\alpha_D+D\mu_{\alpha}\cos\delta_D\cos\alpha_D\nonumber\\
  -D\mu_{\delta}\sin\delta_D\sin\alpha_D\bigr )+\cos\delta_*\cos\alpha_*\bigl(V_D\cos\delta_D\cos\alpha_D\nonumber\\
  -D\mu_{\delta}\sin\delta_D\cos\alpha_D-D\mu_{\alpha}\cos\delta_D\sin\alpha_D\bigr )\nonumber\\
  +\sin\delta_*\bigl(V_D\sin\delta_D+D\mu_{\delta}\cos\delta_D \bigr ).
  \label{eq:vrel}
\end{eqnarray}

\subsection{Mg-Index Distributions}
\label{subsec:mgdist}

We assume that the 1-D distributions of reduced magnesium index $W'$ in metal-rich and metal-poor dSph subcomponents are Gaussian and independent of radius such that, for example, the probability distribution of magnesium indices for the metal-rich subcomponent is
\begin{equation}
  p_{W',1}(W')=\frac{1}{\sqrt{2\pi (\sigma^2_{W',1}+\epsilon^2_{W'})}}\exp\biggl [-\frac{1}{2}\frac{(W'-\langle W'\rangle_{1})^2}{\sigma^2_{W',1}+\epsilon^2_{W'}}\biggr ]
  \label{eq:wdisp}
\end{equation}
(and similarly for the metal-poor subcomponent), where $\langle W'\rangle_1$ is the mean spectral index, $\sigma_{W',1}$ is the intrinsic spectral index dispersion and $\epsilon_{W'}$ is the measurement error.

\subsection{Foreground Distributions}
\label{subsec:foreground}

Compared to dSph member populations, the Milky Way foreground has a more uniform spatial distribution, a wider velocity distribution, and systematically larger magnesium indices.  We quantify each of these distributions directly from the MMFS data using estimates of membership probabilities, $P_{\mathrm{mem}}$, tabulated by W09.  We then smooth the data with Gaussian kernels to estimate probability distributions of radius, velocity and reduced magnesium index for the foreground contamination present in the data:

\begin{eqnarray}
  \hat{p}_{R,\mathrm{MW}}(R)=\frac{\displaystyle\sum_{i=1}^{N_{\mathrm{sample}}}\frac{1-P_{\mathrm{mem},i}}{\sqrt{2\pi k_2^2}}\exp\biggl [-\frac{1}{2}\frac{(R_i-R)^2}{k_2^2} \biggr ]}{\displaystyle\sum_{i=1}^{N_{\mathrm{sample}}}(1-P_{\mathrm{mem},i})};\nonumber\\
  \hat{p}_{V,\mathrm{MW}}(V)=\frac{\displaystyle\sum_{i=1}^{N_{\mathrm{sample}}}\frac{1-P_{\mathrm{mem},i}}{\sqrt{2\pi \epsilon^2_{V,i}}}\exp\biggl [-\frac{1}{2}\frac{(V_i-V)^2}{\epsilon^2_{V,i}} \biggr ]}{\displaystyle\sum_{i=1}^{N_{\mathrm{sample}}}(1-P_{\mathrm{mem},i})};\nonumber\\
  \hat{p}_{W',\mathrm{MW}}(W')=\frac{\displaystyle\sum_{i=1}^{N_{\mathrm{sample}}}\frac{1-P_{\mathrm{mem},i}}{\sqrt{2\pi \epsilon^2_{W',i}}}\exp\biggl [-\frac{1}{2}\frac{(W'_i-W')^2}{\epsilon^2_{W',i}} \biggr ]}{\displaystyle\sum_{i=1}^{N_{\mathrm{sample}}}(1-P_{\mathrm{mem},i})},
   \label{eq:foreground}
\end{eqnarray} 
where $\epsilon_V$ and $\epsilon_{W'}$ are again measurement errors.  In general the distributions of foreground stars do not change significantly over the region ($\sim 1$ square degree) subtended by a dSph, so the only spatial scales built into the data are those corresponding to the dSph structural parameters and observational selection effects (Section \ref{subsec:selection}).  In order to avoid introducing additional scales into our analysis, we set the spatial smoothing bandwidth, $k_2$, equal to the halflight radius measured for the composite dSph stellar component.   We adopt halflight radii measured by \citet{ih95} and tabulated in the erratum to \citet{walker09d}.  For our tests with synthetic data (Section \ref{sec:tests}) we adopt the halflight radius that we estimate from the composite stellar population before considering two-subcomponent models.

\subsection{Likelihood Function Revisited}
\label{simplified}

We have now specified parametric models for the 1D probability distributions of position, velocity and magnesium index for both dSph stellar subcomponents and the Milky Way foreground contamination.  Because our adopted models for velocity and magnesium index distributions are independent of radius, the joint probability distributions are separable into products of the 1-D distributions adopted in Sections \ref{subsec:rdist} - \ref{subsec:foreground}, such that 
\begin{eqnarray}
  p_{1}(R,V,W')=p_{R,1}(R)p_{V,1}(V)p_{W',1}(W');\nonumber\\
  p_{2}(R,V,W')=p_{R,2}(R)p_{V,2}(V)p_{W',2}(W');\nonumber\\
  p_{\mathrm{MW}}(R,V,W')=p_{R,\mathrm{MW}}(R)p_{V,\mathrm{MW}}(V)p_{W',\mathrm{MW}}(W').  
  \label{eq:jointprob}
\end{eqnarray}
After making these substitutions and recognizing that $\int p_{V,1}(V)dV=\int p_{V,2}(V)dV=\int p_{W',1}(W')dW'=\int p_{W',2}(W')dW'=1$ by construction, the likelihood given by Equation \ref{eq:likelihood} becomes
\begin{eqnarray}
  L(\{R_i,V_i,W'_i\}_{i=1}^{N_{\mathrm{sample}}}|\vec{S})=\hspace{1.75in}\nonumber\\
  \displaystyle\prod_{i=1}^{N_{\mathrm{sample}}}\biggl [ f_{1}\frac{w(R_i)p_{R,1}(R_i)p_{V,1}(V_i)p_{W',1}(W'_i)}{\int_{0}^{\infty} w(R)p_{R,1}(R)dR}\nonumber\\
  +f_{2}\frac{w(R_i)p_{R,2}(R_i)p_{V,2}(V_i)p_{W',2}(W'_i)}{\int_{0}^{\infty} w(R)p_{R,2}(R)dR}\nonumber\\
  +(1-f_{1}-f_{2})\hat{p}_{\mathrm{MW},R}(R_i)\hat{p}_{\mathrm{MW},V}(V_i)\hat{p}_{\mathrm{MW},W'}(W'_i)\biggr ].
  \label{eq:likelihood2}
\end{eqnarray}
For the term representing the Milky Way foreground we have substituted the probability distributions estimated by smoothing the data (section \ref{subsec:foreground}), which necessarily include the effects of sampling bias such that $\hat{p}_{\mathrm{MW},R}(R_i)\hat{p}_{\mathrm{MW},V}(V_i)\hat{p}_{\mathrm{MW},W'}(W'_i)\approx w(R_i)p_{\mathrm{MW}}(R_i,V_i,W'_i)\bigl [\int\int\int w(R)p_{\mathrm{MW}}(R,V,W')dRdVdW'\bigr ]^{-1}$.  

\subsection{Markov-Chain Monte Carlo Technique}
\label{subsec:mcmc}

We require a total of twelve free parameters in order to evaluate the likelihood given by Equation \ref{eq:likelihood2}.  Eight parameters specify for both member subcomponents the halflight radii (Equation \ref{eq:plummerprob}), velocity dispersions (Equation \ref{eq:vprob}), and means and variances of the reduced Mg-index distributions (Equation \ref{eq:wdisp}).  Two parameters, $f_{\mathrm{mem}}\equiv (N_{1}+N_{2})/(N_{1}+N_{2}+N_{\mathrm{MW}})$ and $f_{\mathrm{sub}}\equiv (N_{1}/(N_{1}+N_{2})$, specify the fractions $f_{1}=f_{\mathrm{mem}}f_{\mathrm{sub}}$ and $f_{2}=f_{\mathrm{mem}}(1-f_{\mathrm{sub}})$.  The final two parameters, $\mu_{\alpha}$ and $\mu_{\delta}$, specify the two components of the dSph proper motion (Equation \ref{eq:vrel}).  

For all parameters we adopt uniform priors over ranges that include all reasonable values (Table \ref{tab:priors}).  Notice that we specify halflight radii of the stellar subcomponents in terms of free parameters $\log_{10}[r_{h,2}/\mathrm{pc}]$ and the ratio $[r_{h,1}/r_{h,2}]$, and we specify the mean reduced Mg indices in terms of free parameters $\langle W' \rangle_{1}/$\AA\ and the difference $(\langle W'\rangle_{1}-\langle W'\rangle_{2})/$\AA.  This formulation lets us apply priors $0\leq [r_{h,1}/r_{h,2}] \leq 1$ and $0\leq (\langle W'\rangle_{1}-\langle W'\rangle_{2})/$\AA$\leq 3$.  In other words, we assume that the metal-rich subcomponent is more centrally concentrated than the metal-poor subcomponent, as indicated by previous studies of Sculptor \citep{tolstoy04}, Fornax \citep{battaglia06} and Sextans \citep{battaglia11}.

\begin{deluxetable*}{lrrlrrrr}
  \tabletypesize{\scriptsize}
  \tablewidth{0pc}
  \tablecaption{Boundaries of Uniform Priors for Free Parameters in Likelihood Function}
  \tablehead{\\
    \colhead{Parameter}&\colhead{Minimum}&\colhead{Maxium}&\colhead{Description}\\
  }
  \startdata
  $f_{\mathrm{mem}}$&$0$&$1$&$\equiv (N_{1}+N_{2})/(N_{1}+N_{2}+N_{\mathrm{\mathrm{MW}}})$, fraction of stars belonging to dSph\\
  $f_{\mathrm{sub}}$&$0$&$1$&$\equiv N_{1}/(N_{1}+N_{2})$, fraction of members belonging to MR subcomponent\\
  $r_{h,1}/r_{h,2}$&$0$&$1$&ratio of halflight radii for metal-rich (MR) and metal-poor (MP) subcomponents\\
  $\log_{10}[r_{h,2}/\mathrm{pc}]$&$0$&$4.5$&halflight radius of MP subcomponent\\
  $\langle W'\rangle_{1}/$\AA&$-3$&$+3$&mean reduced Mg index of MR subcomponent\\
  $(\langle W'\rangle_{1}-\langle W'\rangle_{2})/$\AA&$0$&$3$&offset of mean Mg indices\\
  $\log_{10}[\sigma^2_{W',1}/$\AA$^2]$&$-5$&$+1$&squared dispersion of reduced Mg index, MR subcomponent\\
  $\log_{10}[\sigma^2_{W',2}/$\AA$^2]$&$-5$&$+1$&squared dispersion of reduced Mg index, MP subcomponent\\
  $\log_{10}[\sigma^2_{V,1}/\mathrm{(km^2s^{-2})}]$&$-5$&$+5$&squared velocity dispersion, MR subcomponent\\
  $\log_{10}[\sigma^2_{V,2}/\mathrm{(km^2s^{-2})}]$&$-5$&$+5$&squared velocity dispersion, MP subcomponent\\
  $\mu_{\alpha}/\mathrm{(mas/century)}$&$-1000$&$+1000$&RA proper motion of dSph\\
  $\mu_{\delta}/\mathrm{(mas/century)}$&$-1000$&$+1000$&Dec. proper motion of dSph\\
  \enddata
  \label{tab:priors}
\end{deluxetable*}

We sample the large parameter space using standard Markov-Chain Monte Carlo techniques.  Specifically we use the Metropolis-Hastings algorithm \citep{metropolis53,hastings70}, which samples the parameter space according to the following prescription: i) from the current location in parameter space $\vec{S}_n$, draw a prospective new location, $\vec{S}'$, from a Gaussian proposal density centered on $\vec{S}_n$; ii) evaluate the ratio of likelihoods at $\vec{S}_n$ and $\vec{S}'$; and iii) if $L(\vec{S}')/L(\vec{S}_n)\ge 1$, accept such that $\vec{S}_{n+1}=\vec{S}'$, else accept with probability $L(\vec{S}')/L(\vec{S}_n)$, such that $\vec{S}_{n+1}=\vec{S}'$ with probability $L(\vec{S}')/L(\vec{S}_n)$ and $\vec{S}_{n+1}=\vec{S}_n$ with probability $1-L(\vec{S}')/L(\vec{S}_n)$.  

We implement the Metropolis-Hastings algorithm using the adaptive MCMC engine CosmoMC\footnote{available at http://cosmologist.info/cosmomc} \citep{lewis02}.  CosmoMC provides a generic sampler that periodically updates the proposal density according to parameter covariances in order to optimize the acceptance rate.  For a given galaxy we run four chains simultaneously, stopping when either the variances of parameter values across the four chains become less than 1\% of the mean of the variances.  Our chains typically require of order $\sim 10^5$ steps to satisfy this convergence criterion.

We process the chain output in two ways.  First, in order to allow for a `burn-in' period during which the chains evolve from initial values to regions of high likelihood, we discard the first half of accepted points.  Second, because the nature of the algorithm introduces correlations between adjacent accepted points, we `thin' the chains by passing only every $25^{\mathrm{th}}$ accepted point to a single `final' chain.  The MCMC method is designed such that points in the final chain randomly sample the posterior probability distribution function (PDF) in the full twelve-dimensional parameter space.  One obtains marginalized probability distributions for any single parameter, or combination of parameters, simply by counting the number of points in the final chain that fall within binned ranges of parameter values \citep{lewis02}. 

The final chains sample, among other distributions, the one-dimensional posterior PDFs of free parameters $\log_{10}[r_{h,2}/\mathrm{pc}]$, $[r_{h,1}/r_{h,2}]$, $\log_{10}[\sigma^2_{V,1}/\mathrm{(km^2s^{-2})}]$ and $\log_{10}[\sigma^2_{V,2}/\mathrm{(km^2s^{-2})}]$.  For every point in the final chains we apply Equation \ref{eq:slope} to obtain the corresponding slope of the mass profile.  The distribution of these slopes then represents our constraint on $\Gamma$.

\section{Tests}
\label{sec:tests}

Our adoption of Plummer profiles and Gaussian distributions to characterize stellar positions, velocities and spectral indices (Section \ref{sec:method}) is motivated by preferences for simplicity and mathematical convenience.  While they are frequently invoked in analyses of dSph data, there is of course no guarantee that these particular models provide accurate descriptions of real dSphs.  In fact, jointly these models can correspond to chemodynamical distribution functions that are unphysical.  

For example, our method twice employs the assumption that both dSph stellar subcomponents have constant velocity dispersion.  This assumption is implicit in the mass estimator given by Equation \ref{eq:walker} (see Section 3.6 of \citealt{walker09d}) and explicit in the models adopted for subcomponent velocity distributions (Equation \ref{eq:vprob}).  While the composite stellar populations of dSphs generally exhibit flat velocity dispersion profiles over the range of radii where data are available, all (Newtonian) equilibrium dynamical models require that tracer velocity dispersion profiles eventually decline at large radius.  Furthermore, even in the regions where data indicate that the composite dSph stellar populations have flat velocity dispersion profiles, there is no guarantee that the velocity dispersion profiles of either/both stellar subcomponents are flat individually.  Indeed, \citet{battaglia08} find that the metal-rich subcomponent of the Sculptor dSph has a falling velocity dispersion profile (see their Figure 3).

We must therefore examine the reliability of our method when it operates on data sets representing realistic dynamical systems that violate our simplistic modeling assumptions.  In order to accomplish this task, we apply our method to a series of synthetic data sets and compare the resulting estimates of stellar population parameters, masses and slopes to known input values.  We construct a given synthetic data set by sampling the superposition of three (two dSph-like member subcomponents plus foreground contamination) chemo-dynamical distribution functions.  The phase-space distribution functions of both member subcomponents correspond to independent (except for the obvious constraint that both have the same  gravitational potential) physical dynamical models in which the stellar subcomponent traces a gravitational potential that is generated by a dark matter halo.  Here we describe these dynamical models, the generation of synthetic data sets, and the systematic behaviors of the errors we identify by applying our method to these synthetic data.

\subsection{Dynamical Models for Individual Stellar Subcomponents} 
\label{subsec:models}

In order to generate test cases we consider physical dynamical models in which two stellar subcomponents independently trace the same dark matter potential.  We build these models simply by combining the two dynamical models that describe the stellar subcomponents individually.  We consider individual stellar subcomponents that are each distributed according to a generalized Hernquist density profile \citep{hernquist90,zhao96},
\begin{equation}
  \nu_{\mathrm{*}}(r)=\nu_{0}\biggl (\frac{r}{r_{\mathrm{*}}}\biggr )^{-\gamma_{\mathrm{*}}} \biggl [1+\bigl (\frac{r}{r_{\mathrm{*}}}\bigr )^{\alpha_{\mathrm{*}}}\biggr ]^{(\gamma_{\mathrm{*}}-\beta_{\mathrm{*}})/\alpha_{\mathrm{*}}},
  \label{eq:nu}
\end{equation}
and we consider dark matter halos with density profiles that take the same form,
\begin{equation}
  \rho_{\mathrm{DM}}(r)=\rho_0\biggl (\frac{r}{r_{\mathrm{DM}}}\biggr )^{-\gamma_{\mathrm{DM}}} \biggl [1+\bigl (\frac{r}{r_{\mathrm{DM}}}\bigr )^{\alpha_{\mathrm{DM}}}\biggr ]^{(\gamma_{\mathrm{DM}}-\beta_{\mathrm{DM}})/\alpha_{\mathrm{DM}}}.
  \label{eq:rho}
\end{equation}
These profiles have independent parameters specifying normalization, scale radius, inner logarithmic slope ($\gamma$, subscripts omitted for brevity), outer logarithmic slope ($\beta$), and the sharpness ($\alpha$) of the transition between the two slopes.   

We aim to test specifically whether our method can distinguish dark matter halos having constant-density cores ($\gamma_{\mathrm{DM}}=0$) from those that have NFW-like cusps ($\gamma_{\mathrm{DM}}=1$).  Therefore we consider dynamical models in which the central slope of the dark matter density profile takes values of either $\gamma_{\mathrm{DM}}=0$ or $\gamma_{\mathrm{DM}}=1$.  We hold fixed other halo parameters at scale radius $r_{\mathrm{DM}}=1$ kpc, outer slope $\beta_{\mathrm{DM}}=3$ and $\alpha_{\mathrm{DM}}=1$.  For the stars, we consider stellar subcomponents that have structural parameters $\alpha_{\mathrm{*,1}}=\alpha_{\mathrm{*,2}}=2$ and $\gamma_{*,1}=\gamma_{*,2}=0.1$, a range of outer slopes $\beta_*=4,5,6$, and a range of scale radii $r_{*}/r_{\mathrm{DM}}=0.1,0.25,0.5,1,1.5$ corresponding to various degrees of `embeddedness' within the dark matter halo.  

In order to generate synthetic data sets we must first calculate phase-space distribution functions for each stellar subcomponent in each dark matter halo.  For this purpose we consider the family of spherical, anisotropic distribution functions discussed by \citet{osipkov79} and \citet{merritt85}.  These models have velocity distributions with anisotropy profiles of the form $\beta_{\mathrm{ani}}(r)\equiv 1-\bar{v_{\theta}^2}/\bar{v_r^2}=r^2/(r^2+r^2_a)$.  We consider values for the anisotropy radius $r_a$ that give the stellar subcomponent a velocity distribution that either is isotropic at all radii ($r_a=\infty$) or gradually changes from isotropic at small radii to radially anisotropic at large radii ($r_a=r_*$).  Having specified the profiles $\nu(r)$, $\rho(r)$ and $\beta_{\mathrm{ani}}(r)$ for each stellar subcomponent in each dark matter halo, we use Equation 11 of \citet{merritt85} to calculate the corresponding phase-space distribution functions.  We check this calculation by performing N-body simulations in which stars orbit within the adopted potential and have initial positions/velocities drawn from the calculated distribution function.  These simulations show no significant departures from the initial dynamical configuration after 100 crossing times, indicating that the calculated distribution functions indeed correspond to equilibrium dynamical models.  

\begin{figure}
  \epsscale{1.}
  \plotone{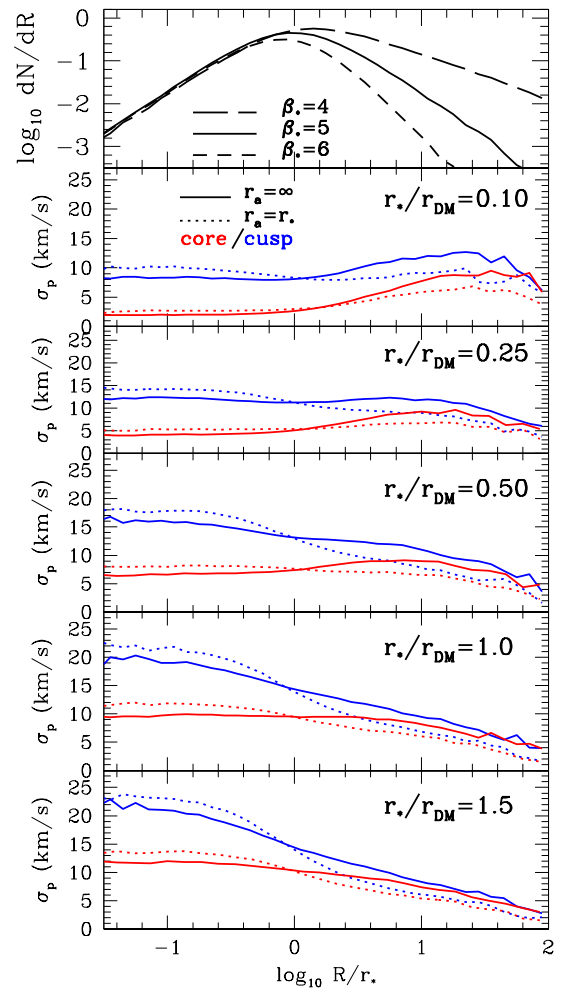}
  \caption{\scriptsize \textit{Top:} Differential distribution of stars as a function of (projected) radius for the stellar subcomponents in our tests.  \textit{Bottom five panels:} Projected velocity dispersion profiles for the physical dynamical models used to construct the synthetic data sets on which we test our method.  For clarity we plot profiles corresponding only to models with outer stellar density profiles specified by $\beta_*=5$ (velocity dispersion profiles for models with $\beta_*=4$ and $\beta_*=6$ behave similarly).  Notice that for a given halo potential, projected velocity dispersion profiles corresponding to isotropic ($r_a=\infty$) and anisotropic ($r_a=r_*$) velocity distributions cross at $r\sim r_*$, the radius where the number of stars reaches a maximum.  This phenomenon helps to explain the insensitivity of our mass estimates to velocity anisotropy (Section \ref{subsubsec:masses}).
}
  \label{fig:prof_vdisp}
\end{figure}

Table \ref{tab:tests} lists the grid of input parameters that specifies 60 unique dynamical models that we use to represent individual dSph stellar subcomponents.  The top panel of Figure \ref{fig:prof_vdisp} displays the differential spatial distributions of stars and illustrates the effect of varying the slopes of the outer stellar density profiles over the range $\beta_*=4,5,6$.  We note that the Plummer surface brightness profiles that we assume in our likelihood function (Section \ref{subsec:rdist}) correspond to (deprojected) stellar density profiles with $(\alpha_*,\beta_*,\gamma_*)=(2,5,0)$.  Test cases with $\beta_*=4$ and $\beta_*=6$ correspond to stellar density profiles that decline less and more steeply, respectively, thereby violating the simplistic assumption of Plummer surface brightness profiles that is inherent in our method.  The bottom five panels in Figure \ref{fig:prof_vdisp} display the projected velocity dispersion profiles that we obtain from large ($N=10^5$) random samples of the corresponding distribution functions.  These test models include velocity dispersion profiles that are flat, rising and/or falling, thereby violating the simplistic assumption of constant velocity dispersion that is inherent in our method.  

\begin{deluxetable}{llllllllrrrrrrrrrrrrr}
  \tabletypesize{\scriptsize}
  \tablewidth{0pc}
  \tablecaption{Tests on synthetic data: grid of input parameters for dynamical test models}
  \tablehead{
    \colhead{Profile}&\colhead{Parameter}&\colhead{values considered}\\}
  \startdata
  Stellar Subcomponent (Eq. \ref{eq:nu})\\
  &$r_*/r_{\mathrm{DM}}$&$0.10,0.25,0.50,1.0,1.5$\\
  &$\alpha_*$&$2$\\
  &$\beta_*$&$4,5,6$\\
  &$\gamma_*$&$0.1$\\
  &$r_a/r_*$&$1,\infty$\\
  Dark Matter Halo (Eq. \ref{eq:rho})\\
  &$\rho_0/(M_{\odot}\mathrm{pc}^{-3})$&$0.064$\\
  &$r_{\mathrm{DM}}/\mathrm{kpc}$&$1$\\
  &$\alpha_{\mathrm{DM}}$&$1$\\
  &$\beta_{\mathrm{DM}}$&$3$\\
  &$\gamma_{\mathrm{DM}}$&$0,1$\\
  \enddata
  \label{tab:tests}
\end{deluxetable}

\subsection{The Meaning of $\Gamma$}
\label{subsec:meaning}

\begin{figure}
  \plotone{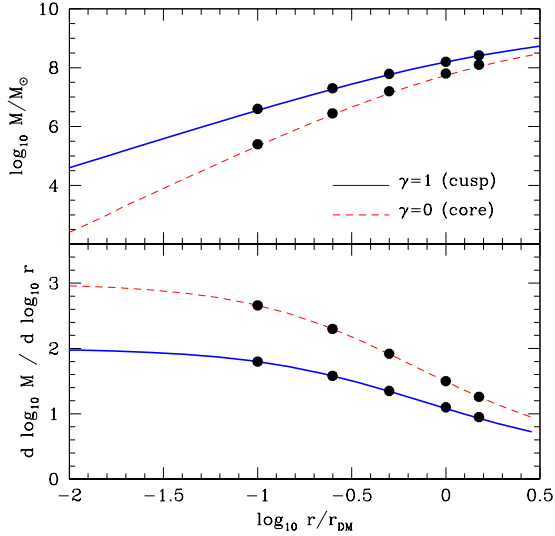}
  \caption{\scriptsize Enclosed mass profiles (top) and slopes of logarithmic mass profiles (bottom), for the cored and cusped dark matter halos considered in tests of our method.  Discrete points identify the luminous scale radii of various dSph-like stellar subcomponents that we embed in these halos for our tests.}  
  \label{fig:prof_slope}
\end{figure}

For the cored and cusped dark matter halos that we consider in our tests, the top panel of Figure \ref{fig:prof_slope} displays enclosed-mass profiles $M(r)$ and the bottom panel displays logarithmic slopes $d\log M/d\log r$.  This figure illustrates the simple relationships between the inner slope of the logarithmic density profile ($\gamma_{\mathrm{DM}}$), the inner slope of the logarithmic mass profile ($\lim_{r\rightarrow 0}[d\log M/d\log r]$), and the slope ($\Gamma\equiv \Delta\log M/\Delta\log r$) that we actually measure.  For the spherically symmetric dark matter density profiles specified by Equation \ref{eq:rho}, the inner mass profile is given by $\lim_{r\rightarrow 0}M(r)\propto r^{3-\gamma_{\mathrm{DM}}}$.  Thus the inner slope of the logarithmic mass profile is given by 
\begin{equation}
  \displaystyle\lim_{r\rightarrow 0}[d\log M/d\log r]=3-\gamma_{\mathrm{DM}}.
  \label{eq:slope2}
\end{equation}
At their centers, cored ($\gamma_{\mathrm{DM}}=0$) dark matter halos have $\lim_{r\rightarrow 0}[d\log M/d\log r]=3$ and cuspy NFW ($\gamma_{\mathrm{DM}}=1$) halos have $\lim_{r\rightarrow 0}[d\log M/d\log r]=2$.  

However, since our method constrains enclosed masses at the halflight radii of two stellar subcomponents, the slope $\Gamma\equiv \Delta\log M/\Delta\log r$ that we measure corresponds to the slope at some finite radius $r> 0$ and will not necessarily represent the central value of $d\log M/d\log r$.  The bottom panel of Figure \ref{fig:prof_slope} illustrates that as $r$ increases, the slope $d\log M/d\log r$ decreases monotonically.  Therefore an unbiased estimate of the slope $\Gamma\equiv \Delta\log M/\Delta\log r$ evaluated at radii $r_{h,2}>r_{h,1}>0$ will necessarily be \textit{smaller} than the central value of $d\log M/d\log r$.  In other words, an unbiased estimate of $\Gamma$ constrains the central value of the slope to be 
\begin{equation}
  \displaystyle\lim_{r\rightarrow 0}[d\log M/d\log r]>\Gamma,
  \label{eq:inequality1}
\end{equation}
and therefore, via Equation \ref{eq:slope2}, constrains the inner slope of the logarithmic density profile to be 
\begin{equation}
  \gamma_{\mathrm{DM}}<3-\Gamma.
  \label{eq:inequality2}
\end{equation}

\subsection{Synthetic Data}
\label{subsec:syntheticdata}

Each of the 60 unique dynamical models included in the grid outlined in Table \ref{tab:tests} describes a single equilibrium stellar component embedded in the potential generated by a dark matter halo.  Therefore each model can correspond to either stellar subcomponent in a system that has two such subcomponents.  We consider all possible combinations of scale radii ($r_{*,1}/r_{\mathrm{DM}}$, $r_{*,2}/r_{\mathrm{DM}}$), outer slopes ($\beta_{*,1}$, $\beta_{*,2}$), and anisotropy radii ($r_{a,1}$, $r_{a,2}$) for which both stellar subcomponents are embedded in the same dark matter halo.  These combinations yield a total of 1080 unique, two-subcomponent structural/dynamical models with which we test our method.  In all cases where the two stellar subcomponents have different scale radii, we assign (see below) reduced magnesium indices such that the subcomponent with smaller $r_*$ is the metal-rich subcomponent, consistent with the phenomenology of well-studied Local Group dSphs \citep{tolstoy04,battaglia06,battaglia11}.  When the two stellar subcomponents have the same scale radii (note that in these cases the slope $\Gamma$ is undefined), we arbitrarily choose one to be metal-rich and the other to be metal-poor.  

For each of the 1080 unique structural/dynamical test models we perform ten realizations.  In setting up a given realization we draw stellar population parameters randomly from uniform distributions within the following limits: 
\begin{itemize}
\item sample sizes $3 \leq \log_{10}[N_{1}+N_{2}+N_{\mathrm{MW}}] \leq 4$ (similar to the available MMFS samples)
\item member fractions $0.4\leq (N_{1}+N_{2})/(N_{1}+N_{2}+N_{\mathrm{MW}}) \leq 0.9$ 
\item subcomponent fractions $0.1\leq N_{1}/(N_{1}+N_{2}) \leq 0.9$ 
\item mean systemic velocities (heliocentric rest frame) $0 \leq \langle V\rangle/(\mathrm{km s^{-1}}) \leq 250$
\item mean spectral index $0.3\leq \langle W'\rangle_1/$\AA$\leq 0.5$ for the `metal-rich' subcomponent
\item mean spectral index separation $0 \leq (\langle W'\rangle_1-\langle W'\rangle_2)/$\AA$\leq 0.25$
\item proper motions $-100 \leq \mu_{\alpha}/(\mathrm{mas/cent})\leq +100$ and $-100 \leq \mu_{\delta}/(\mathrm{mas/cent})\leq +100$.
\end{itemize}
We place half (randomly selected) of the synthetic `dSphs' at the (3D) position of Fornax and the other half at the location of Sculptor (Table \ref{tab:mmfs}).  

With the above stellar population parameters specified for a given realization, we then use an accept/reject algorithm to draw the appropriate numbers of positions and velocities from discrete random samplings of the appropriate 6D distribution function.   We then project the positions and velocities along the line of sight in order to mimic observables $R$ and $V$.  Next we assign reduced Mg indices, $W'$, to each star according to whether it is drawn from the `metal-rich' or `metal-poor' subcomponent.  We assign $W'$ values to the metal-rich and metal-poor member stars by drawing values from Gaussian distributions with variances $\sigma^2_{W',1}=\sigma^2_{W',2}=0.02$ \AA$^{2}$ and means drawn randomly from the ranges specified above.  To the line-of-sight velocities of all member stars we apply redshifts $\langle V\rangle_{\alpha_*,\delta_*}$ appropriate to the synthetic dSph's systemic 3D space motion and line of sight (Equation \ref{eq:vrel}).  Finally, we scatter all velocities and $W'$ values according to actual measurement errors drawn randomly from the MMFS data set (median errors are  $\epsilon_{V}\sim 2$ km s$^{-1}$ and $\epsilon_{W'}\sim 0.01$\AA).  

To stars drawn from a `foreground' contamination component we assign positions drawn randomly from a uniform spatial distribution (within the projected position of the outermost member star) and assign velocities drawn randomly from the Besan{\c c}on model of Milky Way stars (filtered by our photometric criteria for selecting dSph red giants) along the line of sight to the either (chosen randomly in each realization) the Fornax or the Sculptor dSph.  To foreground stars we assign $W'$ values and associated errors drawn directly from measurements of probable ($P_{\mathrm{mem}}<0.1$) foreground stars in the MMFS data set.  

\subsection{Systematic Errors}
\label{subsec:mcmcresults}

We apply our method (including the initial estimation of membership probabilities using the single-component EM algorithm of \citealt{walker09b}; see section \ref{subsec:membership}) to each of the 10800 synthetic data sets (ten realizations for each of the 1080 unique structural/dynamical models).  For our tests we are less interested in distributions of parameter estimates than we are interested in distributions of \textit{errors} $E(\vec{S})\equiv \vec{S}-\vec{S}_{\mathrm{input}}$.  Since it is impractical to examine the error distribution obtained in each individual realization, in what follows we consider error distributions obtained after `stacking' (accomplished by drawing a fixed number of points from the final chains obtained for each of the individual realizations that are to be combined) error distributions corresponding to input models that have various parameters of interest (e.g., $\gamma_{\mathrm{DM}}$, $r_a/r_*$, $r_*/r_{\mathrm{DM}}$) in common.  By combining error distributions in this way, the resulting `composite' error distributions effectively are marginalized over other input parameters (e.g., sample sizes, member fractions, spectral-index parameters, etc.) that could vary from realization to realization.

\begin{figure*}
  \plotone{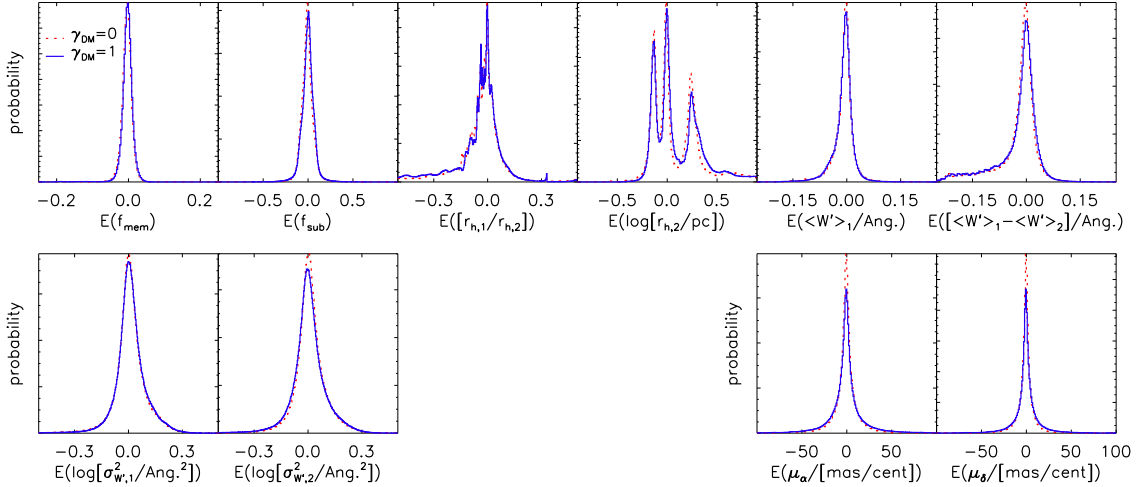}
  \caption{\scriptsize Recovery of the free parameters in our likelihood function (Equations \ref{eq:likelihood} and \ref{eq:likelihood2}), from tests with synthetic data.  Panels display composite error distributions (red/blue for the distributions obtained by stacking error distributions obtained in individual realizations corresponding to cored/cusped input halos) evaluated by subtracting the known input value of each parameter from the MCMC-sampled values.  Peaks at $E(\log_{10}[r_{h,2}/\mathrm{pc}])\sim -0.2, 0.0, +0.3$ correspond to input models with outer stellar density profiles specified by $\beta_{*,2}=6,5,4$, respectively.  Since the test models generally do not have constant velocity dispersion, errors associated with velocity dispersion estimates are poorly defined and therefore not shown (but see Figure \ref{fig:jorge2compnojeansflatpmpmem_masses} for errors associated with derived masses).  }
  \label{fig:jorge2compnojeansflatpmpmem_dmodels}
\end{figure*}

After stacking results according to whether the input dark matter halo is cored or cusped, Figure \ref{fig:jorge2compnojeansflatpmpmem_dmodels} displays error distributions for ten of the twelve free parameters involved in our likelihood function (Equations \ref{eq:likelihood} and \ref{eq:likelihood2}; since the test models do not have constant velocity dispersion, errors associated with free parameters $\sigma^2_1$ and $\sigma^2_2$ are poorly defined and therefore not plotted).  For both cored and cusped input halos, we recover unbiased estimates of most of the free parameters in our likelihood function, as indicated by the symmetry about zero of the composite error distributions shown in Figure \ref{fig:jorge2compnojeansflatpmpmem_dmodels}.  

Notable exceptions appear in panels that display error distributions $E(r_{h,1}/r_{h,2})$ and $E(\log_{10}[r_{h,2}/\mathrm{pc}])$ for the parameters that specify halflight radii of the two stellar subcomponents.  In general we over-/under-estimate halflight radii of subcomponents for which density profiles decline less/more steeply than the Plummer profiles ($\beta_*=5$) adopted in our likelihood function.  For the outer subcomponent, the three peaks at $E(\log_{10}[r_{h,2}/\mathrm{pc}])\sim -0.2$, $0$ and $+0.3$ correspond to tests in which the input model has $\beta_{*,2}=6$, $5$ and $4$, respectively.  Notice that the relative offsets among these peaks directly correspond to offsets in the radii at which the differential spatial distributions $dN/dR$ are maximized for the three stellar density profiles (Figure \ref{fig:prof_vdisp}).  The distribution of errors $E(r_{h,1}/r_{h,2})$ is centered at zero because we tend to recover accurate estimates of $r_{h,1}/r_{h,2}$ for cases in which $\beta_{*,1}=\beta_{*,2}$, even when neither component declines like the assumed Plummer profile.  When the stellar density profile of the inner subcomponent declines less (more) steeply than that of the outer subcomponent, our method tends to over- (under-) estimate the ratio $r_{h,1}/r_{h,2}$, thereby broadening the distribution of errors $E(r_{h,1}/r_{h,2})$ compared to what would be found if we considered only cases with $\beta_{*,1}=\beta_{*,2}$.  Reassuringly, the central peaks at $E(\log_{10}[r_{h,2}/\mathrm{pc}])\sim 0$ and $E(r_{h,1}/r_{h,2}\sim 0$ indicate that we recover unbiased estimates of halflight radii when the input model has the assumed Plummer values of $\beta_{*,1}=\beta_{*,2}=5$.  It is also reassuring that even when our estimates of the subcomponent halflight radii have significant errors, as they tend to when one or both of the stellar subcomponents in the input model have $\beta_*\neq 5$, we recover reasonably unbiased estimates of the other free parameters. 

\subsubsection{Masses}
\label{subsubsec:masses}

As mentioned above, errors associated with free parameters $\sigma^2_1$ and $\sigma^2_2$ are poorly defined because the physical dynamical models that we invoke in order to test our method generally do not have constant velocity dispersion.  It is important to realize that this aspect of the test models violates not only the assumption of constant velocity dispersion (Equation \ref{eq:vprob}) that enters our likelihood function, but also the assumption of constant velocity dispersion that leads to the mass estimator given by Equation \ref{eq:walker} in the first place \citep{walker09d}.  Before we examine the errors associated with the masses returned by our method, let us first examine the bias that is associated directly with the mass estimator on which our method is based.  

\begin{figure}
  \epsscale{1.1}
  \plotone{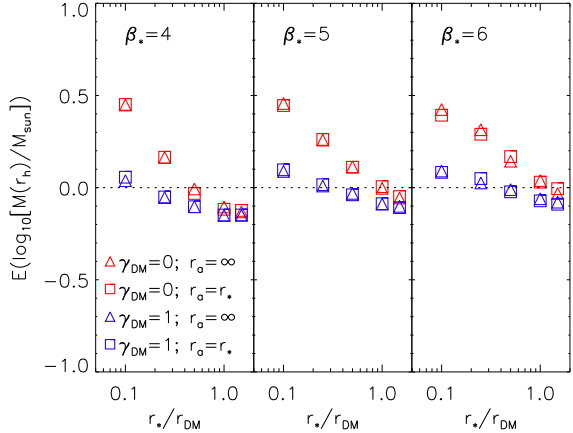}
  \caption{\scriptsize Errors associated with the simple mass estimator given by Equation \ref{eq:walker}, for the dynamical models that we use to construct synthetic data sets.  Models are plotted according to whether the input halo is cored ($\gamma_{\mathrm{DM}}=0$) or cusped ($\gamma_{\mathrm{DM}}=1$), whether the input velocity distribution is isotropic ($r_a=\infty$) or has Osipkov-Merritt anisotropy ($r_a=r_*$), and whether the input stellar density profile has outer slope specified by $\beta_*=4,5,6$.  Regardless of the value of $\gamma_{\mathrm{DM}}$, $r_a$ or $\beta_*$ for the input dynamical model, errors $E(\log_{10}[M(r_{h})/M_{\odot}])\equiv \log_{10}[M(r_h)/M_{\odot}]_{\mathrm{estimated}}-\log_{10}[M(r_h)/M_{\odot}]_{\mathrm{input}}$ correlate primarily with the degree to which the stellar subcomponent is embedded within the dark matter halo.  }
  \label{fig:jorgetest}
\end{figure}

For each of the 60 unique dynamical models (Table \ref{tab:tests}) that we use to represent dSph stellar subcomponents, Figure \ref{fig:jorgetest} plots the difference between the true value of $M(r_{h})$ and the value we obtain by applying our simple mass estimator (Equation \ref{eq:walker}) to an arbitrarily large---and therefore effectively noiseless---sample drawn randomly from the phase-space distribution function.  This figure establishes that there is bias associated directly with our simple mass estimator, and highlights its important characteristics.  

First, the mass errors that are apparent in Figure \ref{fig:jorgetest} show no significant dependence on the velocity anisotropy of the input model.  For a given dark matter halo and embeddedness of the stellar component $r_*/r_{\mathrm{DM}}$, Equation \ref{eq:walker} provides mass estimates having similar errors regardless of whether the tracer velocity distribution is isotropic ($r_a=\infty$) or has a radially variable anisotropy profile ($r_a=r_*$).  This lack of a dependence on anisotropy can be understood phenomenologically upon re-examination of Figure \ref{fig:prof_vdisp}.  There we find that for fixed $\gamma_{\mathrm{DM}}$ and $r_*/r_{\mathrm{DM}}$, projected velocity dispersion profiles corresponding to isotropic and anisotropic distribution functions intersect at approximately the luminous scale radius, $R\sim r_*$.  This is also the radius where the differential spatial distribution $dN/dR$ reaches a maximum (top panel of Figure \ref{fig:prof_vdisp}).  Therefore, samples drawn from both isotropic and the anisotropic distribution functions tend to be dominated by stars at radii where the difference in velocity dispersion between the two cases is negligible.  As a result, we measure similar velocity dispersions and therefore obtain similar mass estimates for isotropic and anisotropic cases.

Second, Figure \ref{fig:jorgetest} shows that while the magnitude of the error depends on whether the dark matter halo is cored or cusped, for both types of halo the error varies with the embeddedness of the stellar component (as quantified by the ratio $r_*/r_{\mathrm{DM}}$) in the same way.  Specifically, values of $M(r_{h}$) obtained from Equation \ref{eq:walker} are more strongly over-estimated for stellar components that are more deeply embedded (smaller $r_*/r_{\mathrm{DM}}$) in their respective dark matter halos.  These tests indicate that this bias is inherent in mass estimators of the form $M(\sim r_{h})\propto r_{h}\sigma^2$ (where $\sigma^2$ is the global velocity dispersion).

\begin{figure}
  \plotone{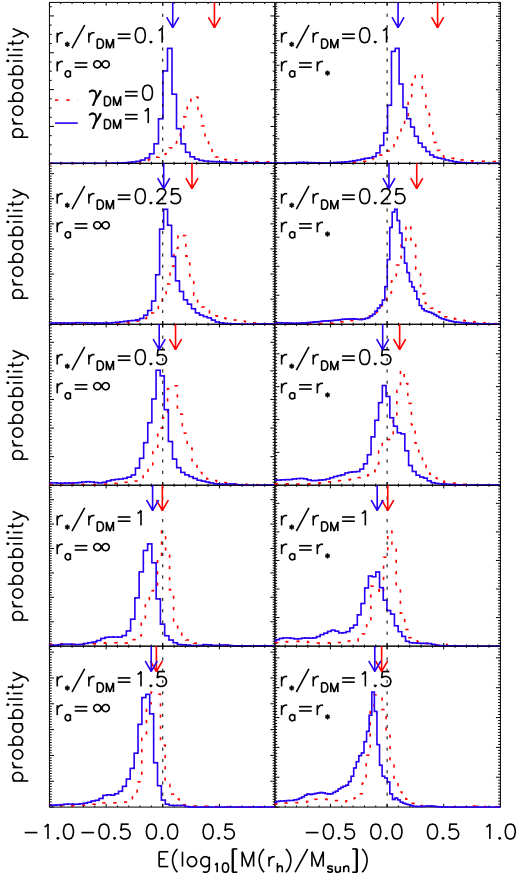}
  \caption{\scriptsize Distributions of mass errors $E(\log_{10}[M(r_{h})/M_{\odot}])\equiv \log_{10}[M(r_{h})/M_{\odot}]-\log_{10}[M_{\mathrm{input}}(r_{h})/M_{\odot}]$ obtained for individual stellar subcomponents in tests of our method using synthetic data (Section \ref{subsubsec:masses}).  Here we have stacked error distributions from individual realizations corresponding to input models that have the same slope for the inner dark matter density profile ($\gamma_{\mathrm{DM}}$), anisotropy radius ($r_a$) and level of embeddedness ($r_*/r_{\mathrm{DM}}$) of the stellar subcomponent within the dark matter halo; we do not separate further according to the outer slope ($\beta_*$) of the stellar density profiles because we find no significant dependence of the mass errors on this quantity.  Downward-pointing arrows identify the mean error that is associated purely with the mass estimator itself (see Figure \ref{fig:jorgetest}).  }
  \label{fig:jorge2compnojeansflatpmpmem_masses}
\end{figure}

Having identified biases inherent in our mass estimator itself, we now examine how those systematic errors propagate through our MCMC analysis.  In order to see how our errors depend not only on the central slope of the dark matter density profile, but also on velocity anisotropy and embeddedness of the stellar subcomponent, we now stack error distributions corresponding to input models that have the same values of $\gamma_{\mathrm{DM}}$, $r_a/r_*$ and $r_*/r_{\mathrm{DM}}$.  We do not break results down further according to the input values of $\beta_{*}$ because we find no significant dependence of the mass errors on this parameter (see, e.g., figure \ref{fig:jorgetest}).  Figure \ref{fig:jorge2compnojeansflatpmpmem_masses} displays the stacked distributions of mass errors $E(\log_{10}[M(r_{h})/M_{\odot}])\equiv \log_{10}[M(r_{h})/M_{\odot}]-\log_{10}[M_{\mathrm{input}}(r_{h})/M_{\odot}]$.  Here we define the mass error as the difference between the estimated mass at the estimated half-light radius and the true mass at the \textit{estimated} half-light radius, since for real dSphs one does not know the subcomponent halflight radii \textit{a priori}.  Figure \ref{fig:jorge2compnojeansflatpmpmem_masses} shows that we generally recover the systematic behavior of errors that we expect to result from the bias inherent in the mass estimator as previously exhibited in Figure \ref{fig:jorgetest} (indicated by downward-pointing arrows in Figure \ref{fig:jorge2compnojeansflatpmpmem_masses}).  More specifically, we recover the expected anti-correlation between $E(\log_{10}[M(r_{h})])$ and $r_{*}/r_{\mathrm{DM}}$.  

We note that for our cored input models with $r_*/r_{\mathrm{DM}}=0.1$ and $r_*/r_{\mathrm{DM}}=0.25$, the distributions of mass errors $E(\log_{10}[M(r_{h})])$ peak at values that are slightly smaller than the errors expected from the bias inherent in the mass estimator (top two rows of panels in Figure \ref{fig:jorge2compnojeansflatpmpmem_masses}).  This peculiarity follows from the fact that for these particular models, the input `global' (i.e., calculated from a large sample of velocities drawn randomly from the subcomponent's distribution function) velocity dispersions are $\sim 2-4$ km s$^{-1}$ (see Figure \ref{fig:prof_vdisp}), similar to the MMFS velocity errors adopted for our tests.  We resolve these small dispersions only marginally and our resulting estimates are biased toward values that are smaller than the input global dispersions.  The corresponding masses are therefore not overestimated as strongly as would be expected from the bias that is inherent in the mass estimator.  This complication does not affect the velocity dispersions (or masses) we measure for subcomponents in real dSphs (Section \ref{sec:results}), which are all $\ga 6$ km s$^{-1}$ and thus well resolved with MMFS data. 

\subsubsection{Slope of the Mass Profile}
\label{subsubsec:slope}

For the present study we are concerned primarily with the error associated with our measurement of the slope $\Gamma$.  The most important result from our tests with synthetic data is the finding that regardless of whether the dark matter halo is cored or cusped, isotropic or anisotropic, our method tends to return more strongly over-estimated masses when the stellar subcomponent is more deeply embedded within the dark matter halo (Figures \ref{fig:jorgetest} and \ref{fig:jorge2compnojeansflatpmpmem_masses}).  Consequently, for a dSph with two stellar subcomponents having different scale radii, our method tends to overestimate the mass of the inner subcomponent more strongly than it does the mass of the outer subcomponent.  Thus we can expect that our estimate of the slope $\Gamma=(\log [M(r_{h,2})/M(r_{h,1})])/(\log [r_{\mathrm{h,2}}/r_{h,1}])$ will tend to be under-estimated.

\begin{figure}
  \plotone{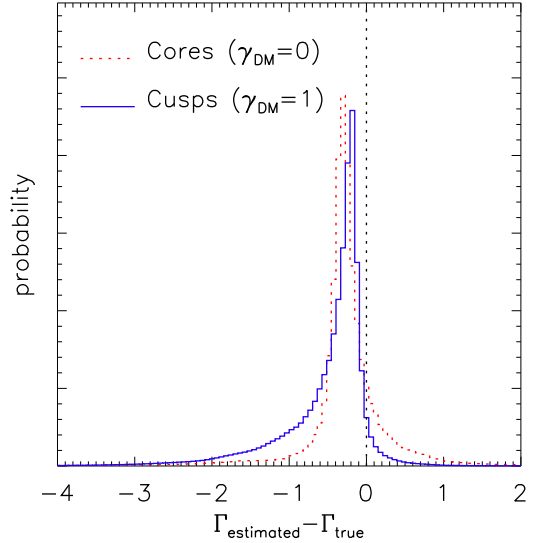}
  \caption{\scriptsize  \textit{Top-Right:} Distributions of slope errors $E(\Gamma)$ obtained in tests of our method using synthetic data (Section \ref{subsubsec:slope}).  Here we have stacked error distributions from individual realizations corresponding to input models that have the same slopes for the inner dark matter density profile ($\gamma_{\mathrm{DM}}$).  We do not include results for test cases in which the input stellar components have the same halflight radii, since the slope $\Gamma$ is then undefined.  In general we identify a bias such that our estimates of the logarithmic slope of the mass profile generally are smaller than known input values (Section \ref{subsubsec:slope}), as expected from the dependence on embeddedness $r_*/r_{\mathrm{DM}}$ of the bias associated with our mass estimator (Figure \ref{fig:jorgetest}).}
  \label{fig:jorge2compnojeansflatpmpmem_gamma}
\end{figure}

This is exactly what we find in our tests.  Figure \ref{fig:jorge2compnojeansflatpmpmem_gamma} displays\footnote{In Figure \ref{fig:jorge2compnojeansflatpmpmem_gamma} we exclude PDFs corresponding to structural/dynamical models in which $r_{*,1}=r_{*,2}$, since for these cases $\Gamma$ is undefined.  Accordingly, the PDFs for $\Gamma$ obtained in these tests indicate no meaningful constraint.} composite distributions of slope errors $E(\Gamma)\equiv\Gamma-(\log [M_{\mathrm{input}}(r_{h,2})/M_{\mathrm{input}}(r_{h,1})])/(\log [r_{\mathrm{h,2}}/r_{h,1}])$, stacked according to the input value of $\gamma_{\mathrm{DM}}$.  As expected, the anti-correlation between $E(\log_{10}[M(r_h)])$ and $r_*/r_{\mathrm{DM}}$ causes the bulk of these distributions to have values $E(\Gamma)<0$, indicating that indeed we tend to underestimate $\Gamma$ for both cored and cusped input halos. 

We note that error distributions obtained in individual realizations of all tested models have finite widths and tails that include some positive errors $E(\Gamma)>0$.  However, of the models that we consider, only those with input values of $\gamma_{\mathrm{DM}}=0$ and $r_{*,2}/r_{\mathrm{DM}}\leq 0.25$ have error distributions that are sometimes \textit{centered}---as quantified by either the maximum- or median-likelihood value---at $E(\Gamma)>0$.  These particular cases are responsible for generating the relatively thick tail toward positive values seen in Figure \ref{fig:jorge2compnojeansflatpmpmem_gamma} for the distribution of $E(\Gamma)$ corresponding to cored input models.  As with the mass errors discussed in the previous section, these peculiarities result from sampling rather than systematic error, since for these particular models both subcomponent velocity dispersions are small ($\sim 2-4$ km s$^{-1}$, see Figure \ref{fig:prof_vdisp}) and only marginally resolved with MMFS-quality data.  Again, this resolution issue does not affect our analysis of real dSphs (Section \ref{sec:results}), for which we measure subcomponent velocity dispersions of $\ga 6$ km s$^{-1}$, at least three times larger than the median MMFS velocity measurement error.

\subsection{Summary of Test Results}
\label{subsec:testconclusions}

We summarize the results of our tests with the following conclusions.  For the ranges of dynamical models and stellar population parameters considered here: 
\begin{itemize}
\item{our method provides reasonably accurate and unbiased estimates of stellar population parameters $f_{\mathrm{mem}}$, $f_{\mathrm{sub}}$, $\langle W'\rangle_1$, $\langle W'\rangle_1-\langle W'\rangle_2$, $\log_{10}[\sigma^2_{W',1}]$ and $\log_{10}[\sigma^2_{W',2}]$ (Figure \ref{fig:jorge2compnojeansflatpmpmem_dmodels}).}
\item{Our method tends to over-/under-estimate halflight radii of stellar subcomponents that have density profiles declining less/more steeply than the Plummer profiles assumed in our likelihood function (Figure \ref{fig:jorge2compnojeansflatpmpmem_dmodels}).}
\item{The assumption of constant velocity dispersion that is inherent in our mass estimator (Equation \ref{eq:walker}) results in bias in our estimates of the masses $M(r_{h})$ of both stellar subcomponents (Figure \ref{fig:jorgetest}).   This bias shows no significant correlation with the velocity anisotropy of the stellar subcomponents (Figures \ref{fig:jorgetest} and \ref{fig:jorge2compnojeansflatpmpmem_masses}).  Rather, this bias correlates primarily with the degree to which the stellar subcomponent is embedded in the dark matter halo, such that (so long as both velocity dispersions are resolved) we tend to overestimate the mass enclosed within the halflight radius of the inner subcomponent more strongly than we do the mass enclosed within the halflight radius of the outer subcomponent (Figures \ref{fig:jorgetest} and \ref{fig:jorge2compnojeansflatpmpmem_masses}).}
\item{This anti-correlation between mass error ($E(\log_{10}[M(r_h)/M_{\odot}])$) and degree of embeddedness ($r_*/r_{\mathrm{DM}}$) of the stellar subcomponent results in a tendency to \textit{under-}estimate the slope of the mass profile for both cored and cusped input halos (Figure \ref{fig:jorge2compnojeansflatpmpmem_gamma}).  }
\end{itemize}
Furthermore, we have already seen (Section \ref{subsec:meaning}) that the value of the slope $\Gamma\equiv \Delta \log M/\Delta\log r$ evaluated at two points $r_{h,2}>r_{h,1}>0$ is smaller than the central value of the slope of the logarithmic mass profile (Figure \ref{fig:prof_slope}).  The fact that we tend to underestimate $\Gamma$ implies that Inequalities \ref{eq:inequality1} and \ref{eq:inequality2} continue to hold if the true value of $\Gamma$ is replaced by our estimate.  

We note that the tests described in this section cover a finite range of dynamical models.  In principle there is no limit to the number and variety of models that might be applied to dSphs.  In addition to the models described in this section, we have performed a similar set of tests using dynamical models that allow for either constant radial ($\beta_{\mathrm{ani}}=+0.25$) or constant tangential ($\beta_{\mathrm{ani}}=-0.45$) velocity anisotropy.  The corresponding distribution functions were sampled by Mark Wilkinson (private communication), as described by \citet{charbonnier11}.  In response to a request from the referee, we have also tested a model that reproduces the velocity dispersion profiles (particularly the steeply falling profile of the inner component) measured by \citet{battaglia08} for metal-rich and metal-poor subcomponents in the Sculptor dSph.  We find that we can reproduce these profiles with a model in which inner (outer) subcomponents follow \citet{king62} surface density profiles with core radii $r_c=195$ pc ($r_c=250$ pc), tidal radii $r_t=6r_c$ ($r_t=15r_c$), Osipkov-Merritt anisotropy radii $r_a=195$ pc ($r_a=500$ pc), and are embedded in a cored ($\gamma_{\mathrm{DM}}=0$) dark matter halo with central density $\rho_0=0.5M_{\odot}/\mathrm{pc}^3$ and scale radius $r_{\mathrm{DM}}=360$ pc.  In our tests of these additional dynamical models we continue to find the same tendency to over-estimate $M(r_h)$ according to the degree to which the stellar component is embedded within the dark matter halo, and thus the same tendency to underestimate $\Gamma$.  

Finally, we note that while the test models that we have considered include complexities (rising/falling velocity dispersion profiles and non-Plummer stellar density profiles) that violate the assumptions of our method and give rise to the systematic errors that we have identified, there is no doubt that real dSphs have additional complexities that we have not simulated in our tests.  Examples may include non-sphericity of stellar and dark matter distributions, non-equilibrium kinematics, internal rotational components, binary stars, the presence of \textit{three} or more distinct stellar subcomponents, etc.   In Section \ref{sec:discussion} we discuss the potential for sensitivity of our results to these possibilities.  

\section{Results for Carina, Fornax and Sculptor}
\label{sec:results}

\begin{figure*}
  \epsscale{1.1}
  \plotone{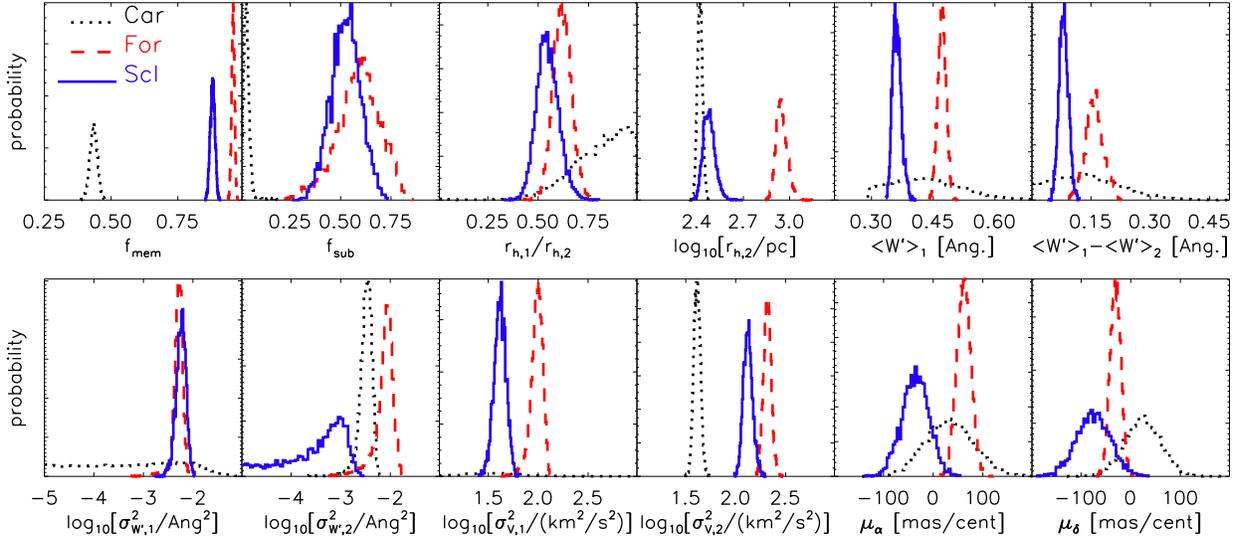}
  \caption{\scriptsize Results for the Carina, Fornax and Sculptor dSphs.   Panels display posterior PDFs for model parameters, obtained from applying the two stellar subcomponent models introduced in Section \ref{sec:method}.  Table \ref{tab:priors} lists median values and $68\%$ ($95\%$) confidence intervals derived from these PDFs.}
  \label{fig:dsph2compnojeansflatpm_params}
\end{figure*}

We now apply our method to the published MMFS data for the Carina, Fornax and Sculptor dSphs.  The top three rows of panels in Figure \ref{fig:dsph2compnojeansflatpm_params} display posterior PDFs for each model parameter.  Table \ref{tab:mcmcresults} lists for each parameter the median value from the posterior PDF, with error bars indicating intervals that enclose the central $68\%$ (and $95\%$) of values.  

\subsection{Carina}
\label{subsec:carina}
Our method returns no compelling evidence for chemo-dynamically distinct stellar subcomponents in Carina, for which we obtain $f_{\mathrm{sub}}= 0.04_{-0.01}^{+0.01}$ (the 95\% error is consistent with $f_{\mathrm{sub}}=0$).  This non-detection may be surprising given Carina's episodic star formation history as evidenced by its two horizontal branches \citep{smeckerhane94} and three main-sequence turn-offs \citep{hurleykeller98}.  However, a previous analysis by \citet{koch06} of an independent VLT/FLAMES data set uncovers only a weak metallicity gradient in Carina, a result qualitatively reproduced using MMFS data \citep{walker09a}.  Thus while Carina may in fact have independent stellar populations, the apparently weak coupling of chemical and dynamical properties prevents a clear separation using our method.  As we do not clearly separate two stellar subcomponents, we obtain no meaningful estimate of $\Gamma$ for Carina.  

For the single stellar component that includes a fraction $f_{\mathrm{mem}}=0.41_{-0.01}^{+0.01}$ of stars in the Carina field, our method returns estimates of the halflight radius ($r_{h}=260_{-10}^{+10}$ pc) and velocity dispersion ($\sigma_V=6.4_{-0.2}^{+0.3}$ km s$^{-1}$) that agree with previously published values based on analyses that assume a single stellar component (e.g., \citealt{mateo93,ih95,munoz06}).  The proper motion returned by our method ($\mu_{\alpha}=+40_{-46}^{+47}$ mas/century, $\mu_{\delta}=+17_{-36}^{+35}$ mas/century; see also \citealt{walker08}) is weakly constrained but agrees with the HST astrometric measurement \citep{piatek03}.  

\subsection{Fornax and Sculptor}
\label{subsec:forscl}

For the remainder of this section we shall consider only Fornax and Sculptor, for both of which our method clearly distinguishes two stellar subcomponents.  We find that a fraction $f_{\mathrm{sub}}= 0.60_{-0.11}^{+0.10}$ of Fornax members belong to a relatively metal-rich subcomponent ($\langle W' \rangle_1=0.47_{-0.01}^{+0.01}$ \AA) with $r_{h,1}=559_{-32}^{+26}$ pc and velocity dispersion $\sigma_{V,1}=10.0_{-0.7}^{+0.6}$ km s$^{-1}$, while the metal-poor ($\langle W' \rangle_2=0.31_{-0.04}^{+0.03}$ \AA) subcomponent has $r_{h,2}=888_{-51}^{+83}$ pc and velocity dispersion $\sigma_{V,2}=14.5_{-0.7}^{+0.6}$ km s$^{-1}$.  For Sculptor we find that a fraction $f_{\mathrm{sub}}=0.53_{-0.8}^{+0.7}$ of members belong to a metal-rich ($\langle W' \rangle_1=0.36_{-0.01}^{+0.01}$ \AA) subcomponent with $r_{h,1}=167_{-10}^{+9}$ pc and velocity dispersion $\sigma_{V,1}=6.5_{-0.5}^{+0.4}$ km s$^{-1}$, while the remaining metal-poor ($\langle W'\rangle_2=0.28_{-0.01}^{+0.01}$ \AA) subcomponent has $r_{h,2}=302_{-24}^{+28}$ pc and velocity dispersion $\sigma_{V,2}=11.6_{-0.6}^{+0.6}$ km s$^{-1}$.  Thus we recover the results of \citet{tolstoy04} and \citet{battaglia06}, who show that the metal-rich inner subcomponents of Sculptor and Fornax have smaller velocity dispersions than the metal-poor outer subcomponents (furthermore, the scale radii of $150\pm 20$ pc and $350 \pm 10$ pc that \citet{battaglia08} derive for inner and outer scale radii, respectively, are broadly consistent with our measurements).  The constraints on Fornax's proper motion stand in excellent agreement with independent astrometric measurements made with HST \citep{piatek07}.  Our measurement of Sculptor's proper motion disagrees with astrometric measurements by \citet{schweitzer95} and \citet{piatek06}, which also disagree with each other.  If either of the astrometric measurements is correct, then the velocity gradient that is present in Sculptor may indicate a small rotational component \citep{battaglia08}.

\begin{figure*}
  \plotone{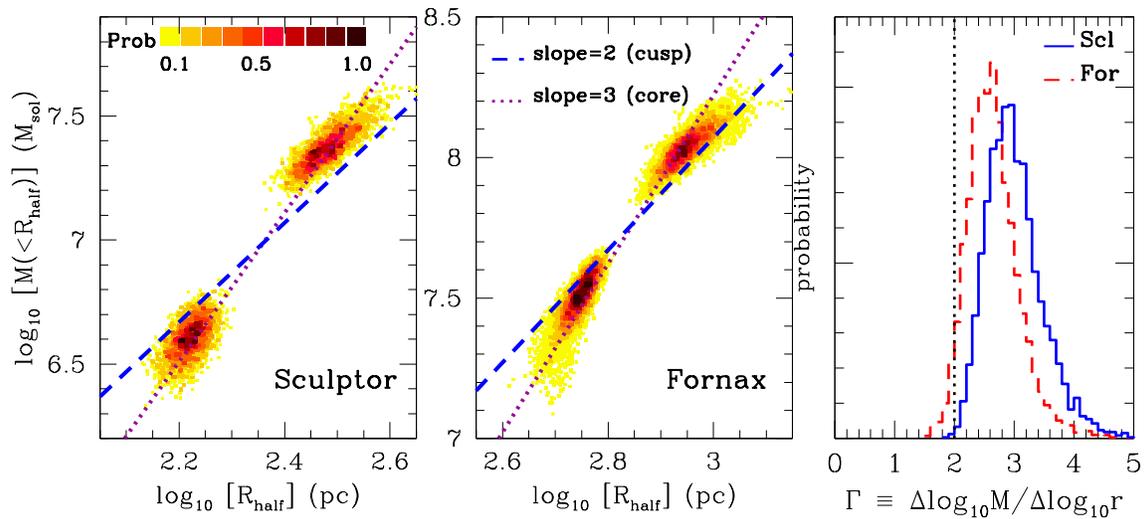}
  \caption{\scriptsize \textit{Left, center:} Constraints on halflight radii and masses enclosed therein, for two independent stellar subcomponents in the Fornax and Sculptor dSphs.  Plotted points come directly from our final MCMC chains, and color indicates relative likelihood (normalized by the maximum-likelihood value).  Overplotted are straight lines indicating the central (and therefore maximum) slopes of cored ($\lim_{r\rightarrow 0}d\log M/d\log r]=3$) and cusped ($\lim_{r\rightarrow 0}d\log M/d\log r]=2$) dark matter halos.  
\textit{Right:} Posterior PDFs for the slope $\Gamma$ obtained for Fornax and Sculptor.  The vertical dotted line marks the maximum (i.e., central) value of an NFW profile (i.e., cusp with $\gamma_{\mathrm{DM}}=1$, $\lim_{r\rightarrow 0}[d\log M/d\log r]=2$).  These measurements rule out NFW and/or steeper cusps ($\gamma_{\mathrm{DM}}\geq 1$) with significance $s\ga 96\%$ (Fornax) and $s\ga 99\%$ (Sculptor).}
  \label{fig:2mass_signif}
\end{figure*}

For both galaxies the estimates of halflight radii and velocity dispersions simultaneously provide estimates of masses $\log_{10}[M_1(r_{h,1})/\mathrm{pc}]$, $\log_{10}[M_2(r_{h,2})/\mathrm{pc}]$ (Equation \ref{eq:walker}), and the slope $\Gamma\equiv \Delta \log M/\Delta \log r$ (Equation \ref{eq:slope}).  Left and center panels in Figure \ref{fig:2mass_signif} display our MCMC-sampled values in the plane of halflight radius and enclosed mass, color-coded according to the likelihood obtained from Equation \ref{eq:likelihood2} (normalized by the maximum-likelihood value).  The right-most panel of Figure \ref{fig:2mass_signif} displays the corresponding posterior PDFs for $\Gamma$ for both Fornax and Sculptor.  Formally, we obtain $\Gamma=2.61_{-0.37}^{+0.43}$ for Fornax and $\Gamma=2.95_{-0.39}^{+0.51}$ for Sculptor.  

\subsection{Significance}

Recall from Section \ref{subsec:meaning} that because we estimate masses at two finite points $r_{h,2}>r_{h,1}>0$, the resulting slope $\Gamma\equiv \Delta\log M/\Delta\log r$ corresponds to a region where the instantaneous slope $d\log M/d\log r$ is smaller than the central value of $3-\gamma_{\mathrm{DM}}$ (Figure \ref{fig:prof_slope}).  Furthermore, the amount by which it is smaller depends on the scale radius of the dark matter halo, a quantity that we do not attempt to constrain here.  As a result, a particular value of $\Gamma$ is strictly inconsistent only with central logarithmic density slopes $\gamma_{\mathrm{DM}}$ that are \textit{larger} than $3-\Gamma$ (Inequality \ref{eq:inequality2}).  This means that measured values in the range $2<\Gamma<3$ are consistent with cores ($\gamma_{\mathrm{DM}}=0$; $d\log M/d\log r<3$ at all nonzero radii), but inconsistent with NFW and steeper cusps ($\gamma_{\mathrm{DM}}\geq 1; d\log M/d\log r < 2$) at all nonzero radii).  

On these grounds we can use the posterior PDFs for $\Gamma$, denoted $P(\Gamma)$, to calculate the significance, $s(\gamma_{\mathrm{DM}})$, with which our measurements exclude dark matter density profiles with central slopes equal to or steeper than a particular value of $\gamma_{\mathrm{DM}}$:
\begin{equation}
  s(\gamma_{\mathrm{DM}})= 1-\frac{\displaystyle\int_{3-\gamma_{\mathrm{DM}}}^{\infty}P(\Gamma)d\Gamma}{\displaystyle\int_{-\infty}^{+\infty}P(\Gamma)d\Gamma}.
  \label{eq:significance}
\end{equation}
Our measurements exclude NFW and steeper cusps at $s(\gamma_{\mathrm{DM}}\geq 1)\geq 95.9\%$ (Fornax) and $s(\gamma_{\mathrm{DM}}\geq 1)\geq 99.8\%$ (Sculptor) significance levels.  
For several reasons we regard these formal exclusion levels as conservative.  First, for the equilibrium dynamical systems that we consider as test cases in Section \ref{sec:tests}, our method tends to overestimate the mass of the inner subcomponent more strongly than it overestimates the mass of the outer subcomponent (Section \ref{subsubsec:masses}), resulting in an underestimate of $\Gamma$ (Section \ref{subsubsec:slope}).  The only counter-examples to this trend that we find in our tests result from poorly resolved velocity dispersions (Sections \ref{subsubsec:masses} - \ref{subsubsec:slope}).  This resolution problem does not affect our results for Fornax and Sculptor, where even the smallest velocity dispersion that we measure ($\sigma_{V,1}=6.5_{-0.5}^{+0.4}$ for Sculptor's inner subcomponent) is several times larger than the median velocity error in the MMFS data set.

Second, by setting the lower limit of the integration in Equation \ref{eq:significance} at the value of the central slope $\lim_{r\rightarrow 0}[d\log M/d\log r]=2$, we extend maximum  generosity to models with $\gamma_{\mathrm{DM}}\geq 1$, which have instantaneous slopes $d\log M/d\log r<2$ at all nonzero radii (Section \ref{subsec:meaning} and Figure \ref{fig:prof_slope}).  At the radii ($\ga 300 \mathrm{pc}$) where we evaluate $\Gamma$ for Fornax and Sculptor, the highest-resolution Aquarius simulations predict that dSph-like CDM halos have $d\log \rho/d\log r \sim 1.3$, or equivalently, $d\log M/d\log r\sim 1.7$ (see Figure 23 of \citealt{springel08}).  Our measurements rule out these slopes with significance $\geq 99.54\%$ (Fornax) and $\geq 99.97\%$ (Sculptor).

Third, we have assumed that the stellar subcomponents contribute negligibly to the gravitational potential.  This assumption generally holds for dSphs, but least so for Fornax, where the dynamical mass-to-light ratio is $M/L_V\sim 10$ in solar units \citep{mateo98}.  If we attempt to remove the stellar contribution to the enclosed mass at each radius using the best-fit Plummer profiles to both stellar subcomponents, we find that for any plausible stellar mass-to-light ratio $0.5\la M/L_V/[M/L_V]_{\odot}\la 5$, our estimates of $\Gamma$ increase by a few percent (because the stars contribute a larger fraction of mass to the inner than to the outer point), again exacerbating the discrepancy with halo models having $\gamma_{\mathrm{DM}}\geq 1$.  In summary, all systematic errors that we have identified behave such that the significance levels we report are conservative.

Finally, we note that for dark matter density profiles of the form given by Equation \ref{eq:rho}, values of $d\log M/d\log r>3$ are unphysical, as they imply $\gamma_{\mathrm{DM}}<0$ (Inequality \ref{eq:inequality2}).  We note that our method does not rule out such unphysical values, which is unsurprising since we have not imposed any physicality constraints.  However, it is reassuring that the bulk of our posterior PDFs correspond to physically plausible scenarios with $\Gamma< 3$.  

\begin{deluxetable*}{llrlrrrr}
  \tabletypesize{\scriptsize}
  \tablewidth{0pc}
  \tablecaption{Constraints from Magellan/MMFS Data: Carina, Fornax and Sculptor\tablenotemark{*}}
  \tablehead{\\
    \colhead{}&\colhead{}&\colhead{Carina}&\colhead{Fornax}&\colhead{Sculptor}&\\
  }
  \startdata
  Model Parameters\\
  \vspace{0.07in}&$f_{\mathrm{mem}}$&$ 0.41_{-0.01(-0.03)}^{+ 0.01(+ 0.03)}$&$ 0.96_{-0.01(-0.01)}^{+ 0.01(+ 0.01)}$&$ 0.88_{-0.01(-0.02)}^{+ 0.01(+ 0.02)}$&\\
  \vspace{0.07in}&$f_{\mathrm{sub}}$&$ 0.04_{-0.01(-0.04)}^{+ 0.01(+ 0.03)}$&$ 0.60_{-0.11(-0.25)}^{+ 0.10(+ 0.18)}$&$ 0.53_{-0.08(-0.15)}^{+ 0.07(+ 0.14)}$&\\
  \vspace{0.07in}&$r_{\mathrm{h,1}}/r_{\mathrm{h,2}}$&$ 0.83_{-0.16(-0.36)}^{+ 0.12(+ 0.17)}$&$ 0.62_{-0.04(-0.09)}^{+ 0.05(+ 0.09)}$&$ 0.55_{-0.05(-0.10)}^{+ 0.06(+ 0.12)}$&\\
  \vspace{0.07in}&$\log_{10}[r_{\mathrm{h,2}}/\mathrm{pc}]$&$ 2.42_{-0.02(-0.03)}^{+ 0.02(+ 0.03)}$&$ 2.95_{-0.03(-0.06)}^{+ 0.04(+ 0.08)}$&$ 2.48_{-0.03(-0.06)}^{+ 0.04(+ 0.09)}$&\\
  \vspace{0.07in}&$\langle W\rangle_{\mathrm{1}}/$\AA&$ 0.44_{-0.08(-0.13)}^{+ 0.09(+ 0.20)}$&$ 0.47_{-0.01(-0.02)}^{+ 0.01(+ 0.02)}$&$ 0.36_{-0.01(-0.02)}^{+ 0.01(+ 0.02)}$&\\
  \vspace{0.07in}&$(\langle W\rangle_{\mathrm{1}}-\langle W\rangle_{\mathrm{2}})/$\AA&$ 0.14_{-0.08(-0.13)}^{+ 0.09(+ 0.20)}$&$ 0.16_{-0.02(-0.05)}^{+ 0.02(+ 0.05)}$&$ 0.08_{-0.01(-0.02)}^{+ 0.01(+ 0.02)}$&\\
  \vspace{0.07in}&$\log_{10}[\sigma^2_{W,\mathrm{1}}/$\AA$^2]$&$-3.15_{-1.25(-1.75)}^{+ 1.10(+ 1.68)}$&$-2.27_{-0.09(-0.33)}^{+ 0.07(+ 0.14)}$&$-2.23_{-0.11(-0.23)}^{+ 0.09(+ 0.18)}$&\\
  \vspace{0.07in}&$\log_{10}[\sigma^2_{W,\mathrm{2}}/$\AA$^2]$&$-2.48_{-0.12(-0.29)}^{+ 0.10(+ 0.18)}$&$-2.07_{-0.19(-0.63)}^{+ 0.12(+ 0.24)}$&$-3.30_{-0.90(-1.58)}^{+ 0.37(+ 0.56)}$&\\
  \vspace{0.07in}&$\log_{10}[\sigma^2_{V,\mathrm{1}}/\mathrm{(km^2s^{-2})}]$&$ 4.46_{-0.28(-5.00)}^{+ 0.19(+ 0.38)}$&$ 2.00_{-0.06(-0.16)}^{+ 0.05(+ 0.09)}$&$ 1.62_{-0.06(-0.13)}^{+ 0.06(+ 0.11)}$&\\
  \vspace{0.07in}&$\log_{10}[\sigma^2_{V,\mathrm{2}}/\mathrm{(km^2s^{-2})}]$&$ 1.62_{-0.03(-0.06)}^{+ 0.03(+ 0.07)}$&$ 2.32_{-0.04(-0.07)}^{+ 0.04(+ 0.08)}$&$ 2.13_{-0.04(-0.08)}^{+ 0.05(+ 0.10)}$&\\
  \vspace{0.07in}&$\mu_{\alpha}/\mathrm{(mas/century)}$&$  40_{ -46( -92)}^{+  47(+  92)}$&$  63_{ -14( -28)}^{+  14(+  27)}$&$ -33_{ -27( -54)}^{+  26(+  51)}$&\\
  \vspace{0.07in}&$\mu_{\delta}/\mathrm{(mas/century)}$&$  17_{ -36( -69)}^{+  35(+  70)}$&$ -29_{ -10( -21)}^{+  10(+  20)}$&$ -72_{ -33( -67)}^{+  34(+  67)}$&\\
  Derived Quantities\\
  \vspace{0.07in}&$\log_{10}[M(r_{\mathrm{h,1}})/M_{\odot}]$& \nodata&$ 7.67_{-0.08(-0.20)}^{+ 0.07(+ 0.12)}$&$ 6.77_{-0.07(-0.15)}^{+ 0.07(+ 0.13)}$&\\
  \vspace{0.07in}&$\log_{10}[M(r_{\mathrm{h,2}})/M_{\odot}]$&$ 6.97_{-0.04(-0.07)}^{+ 0.04(+ 0.07)}$&$ 8.20_{-0.06(-0.12)}^{+ 0.06(+ 0.13)}$&$ 7.53_{-0.07(-0.13)}^{+ 0.08(+ 0.17)}$&\\
  \vspace{0.07in}&$\Gamma\equiv\Delta\log M/\Delta\log r$& \nodata&$    2.61_{   -0.37(   -0.68)}^{+    0.43(+    1.07)}$&$    2.95_{   -0.39(   -0.70)}^{+    0.51(+    1.22)}$&\\
  \vspace{0.07in}&$3-\Gamma$ \tablenotemark{**}&   \nodata&$    0.39_{   -0.43(   -1.07)}^{+    0.37(+    0.68)}$&$    0.05_{   -0.51(   -1.22)}^{+    0.39(+    0.70)}$&\\
  \enddata
  \tablenotetext{*}{Error bars enclose the central $68\%$ ($95\%$) of area under the marginalized 1D posterior probability distribution function.}
  \tablenotetext{**}{From Inequality \ref{eq:inequality2}, this quantity represents an upper limit on the inner slope $\gamma_{\mathrm{DM}}$ of the logarithmic density profile.}
  \label{tab:mcmcresults}
\end{deluxetable*}

\section{Discussion}
\label{sec:discussion}

Let us review the assumptions that enter into our measurement of $\Gamma$.  In formulating our method we assume that a dSph consists of either one or two spherically symmetric, equilibrium stellar subcomponents that independently trace the same spherical dark matter potential.  In order to quantify probability distributions for observed quantities, we further assume that both stellar subcomponents have Plummer surface brightness profiles, Gaussian Mg-index distributions, and Gaussian line-of-sight velocity distributions with constant dispersions that receive negligible contributions from `non-thermal' phenomena such as rotational support and/or binary-orbital motions.  The tests described in Section \ref{sec:tests} indicate that for a range of models that explicitly violate our assumptions about Plummer surface brightness and constant velocity dispersion profiles, our method tends to underestimate $\Gamma$, implying that the stated NFW exclusion limits are conservative.  Here we discuss the potential for sensitivity to several assumptions inherent in our method that are \textit{not} violated in the tests of Section \ref{sec:tests} but might be violated by real dSphs.

\subsection{Spherical symmetry}  
\label{subsec:spherical}

Fornax and Sculptor both have projected minor-to-major axis ratios of $\sim 0.7$ \citep{ih95} and are among the roundest of the Milky Way's dSph satellites.  In order to investigate the degree to which the observed flattening of Fornax and Sculptor might affect our measurements of $\Gamma$, we repeated our analysis using elliptical instead of circular radii, where a star's `elliptical radius' is the semi-major axis of the ellipse (with center listed in Table \ref{tab:mmfs}, position angle and ellipticity listed in Table 2 of \citealt{ih95}) that passes through the position of the star.  Use of elliptical instead of circular radii gives constraints of $\Gamma=2.72_{-0.43}^{+0.50}$ for Fornax (exclusion significance $s(\gamma_{\mathrm{DM}}\geq 1)\geq 96.1\%$) and $\Gamma=2.40_{-0.26}^{+0.32}$ for Sculptor (exclusion significance $s(\gamma_{\mathrm{DM}}\geq 1)\geq 93.9\%$).  Thus the NFW exclusion level for Fornax is relatively robust while the exclusion level for Sculptor shows mild sensitivity to whether or not we adjust for Sculptor's elliptical morphology.

\subsection{Dynamic equilibrium}  
\label{subsubsec:equilibrium}

We measure $\Gamma$ by twice applying a mass estimator (Equation \ref{eq:walker}) derived from the spherical Jeans equation (Equation \ref{eq:jeans}), which holds for spherical systems in dynamic equilibrium.  We must therefore consider the fact that nearby dSphs are vulnerable to disruptive tidal forces as they orbit within the potential of the Milky Way.  Tides can temporarily inflate dSph velocity dispersions---and thus dynamical masses---by `heating' the stellar component during a close pericentric passage and/or by stripping stars that, thence unbound, linger sufficiently near the dSph to be observed and counted as bound members (e.g., \citealt{pp95,oh95,read06,klimentowski07,penarrubia08b,penarrubia09}).  Both phenomena affect the outer more than the inner parts of a satellite---thus tidal heating is the only process we identify that may cause our method to return an \textit{over}-estimate of $\Gamma$.  

However, measurements of their systemic distances and velocities imply that neither Fornax ($D\sim 138$ kpc, \citealt{mateo98}) nor Sculptor ($D\sim 79$ kpc) experience strong tidal encounters with the Milky Way.  Fornax's line-of-sight velocity and proper motion (\citealt{piatek07}, supported by this work) imply a pericenter distance of $r_p=118_{-52}^{+19}$ kpc (\citealt{piatek07}, error bars give 95\% confidence intervals), and Sculptor's imply $r_p\sim 65$ kpc (with 95\% confidence intervals allowing values as low as $\sim 30$ kpc) for either of the two astrometric proper motion measurements \citep{schweitzer95,piatek06}.  N-body simulations by \citet{penarrubia09} and \citet{penarrubia10} demonstrate that for satellite halos that follow the generic density profile given by Equation \ref{eq:rho}, the instantaneous tidal radius at pericenter is $r_t\approx r_p[M_{\mathrm{dsph}}(\leq r_t)/(3M_{\mathrm{MW}}(\leq r_p)]^{1/3}$, where $M_{\mathrm{dsph}}(r_t)$ is the dSph mass enclosed within the tidal radius and $M_{\mathrm{MW}}(\leq r_p)$ is the enclosed mass of the Milky Way within the pericentric distance.  \citet{watkins10} have recently used a sample of tracers (halo stars, globular clusters and satellite galaxies) in the outer Galactic halo to estimate a mass of $M_{\mathrm{MW}}(\leq 300\mathrm{kpc})=0.9\pm 0.3 \times 10^{12}M_{\odot}$.  We obtain conservative lower limits for the pericentric tidal radii of Fornax and Sculptor by considering only the \textit{stellar} mass of each dSph (dynamical masses are $\ga 10$ times larger, \citealt{mateo98}) and taking $M_{\mathrm{MW}}(\leq 300\mathrm{kpc})$ as an upper limit for the mass enclosed within the pericentric radius of each satellite: $r_t\ga r_p[L_{\mathrm{dsph}}\Upsilon_{*}/(3M_{\mathrm{MW}}(\leq 300\mathrm{kpc}))]^{1/3}$, where $\Upsilon_{*}$ is the stellar mass-to-light ratio.  Using the pericentric distances quoted above, the luminosities listed in Table \ref{tab:mmfs} and assuming $\Upsilon_{*}=1 M_{\odot}/L_{\odot}$, we obtain $r_t\ga 2$ kpc for Fornax and $r_t\ga 500$ pc for Sculptor.  These lower limits are larger than the halflight radii that we measure for the outer subcomponents of both galaxies.  Nearly all member stars lie inside these radii and thus---particularly if both galaxies additionally have dark matter halos---we can expect that tides do not profoundly alter the structure and kinematics of Fornax and Sculptor during their pericentric passages.  

Even if Fornax and Sculptor have orbital pericenters much smaller than indicated by their measured proper motions, both systems are now sufficiently far from the Milky Way that they will have reached new equilibrium configurations and shed any stripped stars.  \citet{penarrubia09} demonstrate that unbound tidal debris lingers near a tidally stripped satellite only for a time similar to the satellite's internal crossing time, which for the Milky Way's `classical' dSphs is $t_{c}\sim (300\mathrm{pc})/(10\mathrm{kms^{-1}})\approx 30$ Myr, or the amount of time required to travel $10$ kpc at constant speed $300$ km s$^{-1}$.  This duration is much smaller than the amount of time required by either Fornax or Sculptor to travel to their present locations from pericentric distances compatible with significant tidal disruption.   

\subsection{Rotation}
\label{subsec:rotation}

Mass estimates for stellar subcomponents identified by our method are directly proportional to the corresponding estimates of stellar velocity dispersions.  In principle, any contribution to these velocity dispersions by `non-thermal' motions such as rotational support or unresolved binary orbital motions (next section) might introduce a bias in our mass estimates beyond those that we have already identified in Section \ref{subsubsec:masses}.  

A stellar subcomponent that receives significant support against gravity from rotation about an axis not aligned with the line of sight will exhibit a smooth variation in mean velocity as a function of position.  For the simplest (solid body) rotation models, rotation introduces a gradient in mean line-of-sight velocity.  All three of the dSphs studied here exhibit statistically significant gradients in their velocities as measured in the heliocentric and Milky Way rest frames \citep{walker08,battaglia08}.  However, our method attributes any such gradient not to rotation (which we implicitly assume is insignificant), but wholly to the perspective effect induced by the dSph's systemic motion transverse to the line of sight (Section \ref{subsec:veldist}).  Since we account for this effect in our likelihood function, our method effectively removes the contribution of any apparent velocity gradient from our estimates of the subcomponent velocity dispersions.  

One might object that such gradients can arise due to a combination of perspective effects \textit{and} rotation, and that by attributing any detected gradients entirely to perspective effects, we unduly ignore what might be real and dynamically significant rotation.  This concern is particularly relevant for Sculptor, where the proper motion that we estimate disagrees with both published astrometric measurements \citep{schweitzer95,piatek06}.  However, we find that the gradients in question are sufficiently small that our estimates of $\Gamma$ are insensitive to our assumptions about the relative importance of rotation and perspective effects.  If we assume that Fornax and Sculptor have zero proper motion in the heliocentric rest frame, such that any velocity gradient is allowed to contribute maximally to our velocity dispersion estimates, then we estimate $\Gamma=2.53_{-0.36}^{+0.41}$ for Fornax and $\Gamma=2.94_{-0.39}^{+0.51}$ for Sculptor; both values are statistically indistinguishable from our original constraints.    

\subsection{Binary Stars}
\label{subsec:binaries}

It has long been recognized that dSph mass estimates derived from stellar velocity dispersions may be vulnerable to the inflation of those dispersions by stellar binary orbital motions.  \citet{edo96} and \citet{hargreaves96b} independently demonstrate that unless dSphs have binary orbital distributions that are pathologically different from those observed among field stars, binary motions do not contribute significantly to the velocity dispersions of $\sim 10$ km s$^{-1}$ measured from red giants in `classical' dSphs.  On the other hand, \citet{mcconnachie10} have recently shown that binaries may contribute significantly to (and in some cases dominate) the velocity dispersions of $\la 3-4$ km s$^{-1}$ that have been measured from fainter stars in some of the least luminous Milky Way satellites (e.g., \citealt{martin07,simon07,aden09,koposov11}).  

The Monte Carlo simulations of \citet{mcconnachie10} show that susceptibility to inflation by binary motions increases as the measured velocity dispersion decreases---i.e., for smaller dispersions it becomes more likely that binary motions contribute significantly to the signal.  According to these simulations (and independent simulations by \citealt{minor10}) it is highly unlikely that binary motions contribute significantly to the velocity dispersions we estimate in this study, since the smallest dispersion we estimate is $\sigma_V=6.5_{-0.5}^{+0.4}$ km s$^{-1}$ (for Sculptor's inner subcomponent).  Furthermore, even if binary motions are significant in our sample, they would contribute more significantly to the smaller dispersions that we measure for the inner subcomponents of Fornax and Sculptor than they do to the larger dispersions that we measure for outer subcomponents.  Thus binaries would affect our measurement of the slope $\Gamma$ in the same sense as the bias that is already inherent in our mass estimator (Section \ref{subsubsec:masses}), causing us to overestimate the mass of the inner component to a larger degree than we overestimate the mass of the outer component.  Therefore, to the extent that binary motions are relevant at all, they join the other sources of error that cause us to underestimate $\Gamma$ systematically, again implying that our formal exclusion levels are conservative.  

\subsection{Number of stellar subcomponents}  
\label{subsec:numcomp}

Our method allows for up to two stellar subcomponents.  Deep photometric surveys suggest that Sculptor probably has exactly two stellar subcomponents that are sufficiently bright to observe in large numbers (Sculptor also has extremely metal-poor stars similar to those found in the faintest dSphs \citep{kirby09,frebel10,tafelmeyer10}, which may correspond to a third, `ultrafaint' subcomponent), as indicated by the different spatial distributions of its blue- and red-horizontal-branch stars \citep{hurleykeller99,majewski99}.  Fornax exhibits the same sort of bimodal horizontal branch morphology, but while Sculptor stopped forming stars as long as $\sim 10$ Gyr ago, Fornax has a young main sequence population with age of order $\sim 100$ Myr \citep{stetson98,battaglia06}.  The young main sequence stars are more centrally concentrated than either of the subcomponents traced by evolved giants, so Fornax appears to have at least three distinct stellar subcomponents \citep{battaglia06}.  While the presence of three subcomponents offers the possibility of measuring the slope of Fornax's mass profile using three resolved points, it is unlikely that our spectroscopic sample of stars selected from Fornax's red giant branch includes a significant number from the young and relatively unevolved population.  

\subsection{Future Work}

Finally, we note that the Local Group hosts several more dSphs for which similar measurements are feasible in the short term.  Although its large angular size presents observational challenges, there is already spectroscopic evidence that the Sextans dSph contains chemo-dynamically distinct stellar subcomponents \citep{battaglia11}.  Furthermore, star formation histories measured by comparing observed and synthetic color magnitude diagrams indicate that the Milky Way dSphs Draco, Leo I, Leo II and Ursa Minor all have intermediate ($1\la \mathrm{age/Gyr}\la 8$) as well as old ($\mathrm{age}\ga 10$ Gyr) stellar populations \citep{dolphin05,tolstoy09}.  \citet{harbeck01} identify spatially segregated blue- and red-horizontal branch populations in Andromeda satellites And I and And VI, and \citet{mcconnachie07} find similar photometric evidence for distinct stellar subcomponents in And II.  The ability to separate these objects statistically into chemo-dynamically distinct stellar subcomponents will depend on a combination of spectroscopic sample size and the individual properties of the dSphs, particularly regarding the inherent contrast between stellar subcomponents.  A large and comprehensive survey of dSphs that are separable into stellar subcomponents will help to evaluate the generality of our results for Fornax and Sculptor.

\section{Summary}
\label{sec:summary}

We have introduced a method for measuring the slopes of mass profiles for dSphs that have chemo-dynamically distinct stellar subcomponents.  The method operates directly on spectroscopic data and invokes neither a dark matter halo model nor any assumption about velocity anisotropy.  Rather, it exploits the basic principle that two resolved points in the same profile are sufficient to define a slope.  Given measurements of stellar positions, velocities and spectral indices, our method can measure halflight radii and velocity dispersions simultaneously for up to two stellar subcomponents in the same dSph.  For any mass estimator of the form $M(\kappa r_h)\propto r_h\sigma^2_V$, where $\kappa$ is some constant, these measurements immediately provide an estimate of the slope $\Gamma\equiv \Delta \log M/\Delta\log r$ as defined by two points.  

Applying our method to published Magellan/MMFS data, we distinguish two stellar subcomponents each in the Fornax and Sculptor dSphs, for which we measure $\Gamma=2.61_{-0.37}^{+0.43}$ and $\Gamma=2.95_{-0.39}^{+0.51}$ respectively.  These values are consistent with cored dark matter halos of constant density over the central few hundred parsecs and rule out NFW-like ($d\log M/d\log r \leq 2$) cusps with significance $\ga 96\%$ and $\ga 99\%$, respectively.  Tests with synthetic data drawn from physical distribution functions demonstrate that these exclusion levels are conservative.  These results provide direct evidence against the notion that the current CDM paradigm successfully accounts for the phenomenology of dark matter at small scales.



\section*{Acknowledgments}

MGW's debts to Mario Mateo and Ed Olszewski are immeasurable.  On behalf of the Dwarf Investigative Network (DwaIN), we are grateful to Mark Wilkinson for providing synthetic data sets that we used to perform initial tests of our method.  We thank Gerry Gilmore and colleagues at the Institute of Astronomy (IoA), Cambridge, for support and inspiration.  We thank Alan McConnachie, Justin Read, George Lake and John Magorrian for helpful discussions and comments.  We thank Bodhisattva Sen for guidance regarding statistical issues.  We thank the anonymous referee for suggestions that helped to improve this manuscript.  We thank the IoA for inviting us to teach a graduate seminar on \textit{Dwarf Galaxies and Cosmology} during Winter 2010, when the ideas developed here were formed.  MGW and JP acknowledge support from the STFC-funded Galaxy Formation and Evolution programme at the IoA.  MGW is currently supported by NASA through Hubble Fellowship grant HST-HF-51283.01-A, awarded by the Space Telescope Science Institute, which is operated by the Association of Universities for Research in Astronomy, Inc., for NASA, under contract NAS5-26555.  

%

\end{document}